\documentclass[sigconf]{acmart}

\renewcommand\footnotetextcopyrightpermission[1]{} %
\setcopyright{none}
\acmConference{Review version}{}{}
\copyrightyear{2026}
\acmYear{2026}

\usepackage{amsmath}
\usepackage{amssymb,amsfonts}
\usepackage[linesnumbered,ruled,vlined]{algorithm2e}
\usepackage{graphicx}
\usepackage{textcomp}
\usepackage{xspace}
\usepackage{cleveref}
\usepackage{float}
\usepackage{wrapfig}
\usepackage{subcaption}
\usepackage{multicol}
\usepackage{multirow}
\usepackage{siunitx}
\usepackage{verbatim}
\usepackage{tikz}
\usepackage{tabularx}
\usepackage{enumitem}

\begin{document}

\crefformat{section}{\S~#2#1#3}
\crefformat{subsection}{\S~#2#1#3}
\crefformat{subsubsection}{\S~#2#1#3}
\newcommand\mycommfont[1]{\scriptsize\ttfamily\textcolor{blue}{#1}}
\SetCommentSty{mycommfont}

\newcommand{\AM}[1]{\textcolor{blue}{{\it [AM: #1]}}}
\newcommand{\MMR}[1]{\textcolor{red}{{\it {[MMR: #1]}}}}
\newcommand{\BN}[1]{\textcolor{brown}{{\it {[BN: #1]}}}}
\newcommand{\JM}[1]{\textcolor{olive}{{\it {[JM: #1]}}}}

\newcommand{\proj}{\emph{KVStream}\xspace}
\newcommand{\jlse}{\mbox{Testbed-1}\xspace}
\newcommand{\polaris}{\mbox{Testbed-2}\xspace}

\newcommand{\Tret}{T^{\mathrm{ret}}}
\newcommand{\Tfwd}{T^{\mathrm{fwd}}}
\newcommand{\Twr}{T^{\mathrm{wr}}}
\newcommand{\bbus}{b_{\mathrm{D2H}}}

\title{\proj: Adaptive Parallel Storage Management for KV Caching in Large-Scale LLM Serving}

\author{Avinash Maurya}
\affiliation{
    \institution{Argonne National Laboratory}
    \city{Lemont, IL}
    \country{USA}
}
\email{amaurya@anl.gov}

\author{Jaiaid Mobin}
\affiliation{
    \institution{Rochester Institute of Technology}
    \city{Rochester, NY}
    \country{USA}
}
\email{jm5071@rit.edu}

\author{Jie Ye}
\affiliation{
    \institution{Illinois Institute of Technology}
    \city{Chicago, IL}
    \country{USA}
}
\email{jye20@hawk.illinoistech.edu}

\author{M. Mustafa Rafique}
\affiliation{
    \institution{Rochester Institute of Technology}
    \city{Rochester, NY}
    \country{USA}
}
\email{mrafique@cs.rit.edu}

\author{Franck Cappello}
\affiliation{
    \institution{Argonne National Laboratory}
    \city{Lemont, IL}
    \country{USA}
}
\email{cappello@mcs.anl.gov}

\author{Bogdan Nicolae}
\affiliation{
    \institution{Argonne National Laboratory}
    \city{Lemont, IL}
    \country{USA}
}
\email{bogdan.nicolae@acm.org}
\renewcommand{\shortauthors}{Avinash Maurya et al.}

\begin{abstract}
Long-context LLM serving for agentic, coding, and document workloads reuses prefixes whose key-value (KV) cache outgrows GPU and host memory, putting storage I/O on the critical path of the inference step for retrieving prefixes. A single node-local NVMe absorbs all retrievals, while the parallel file system or cloud file store that nodes already mount carries almost no inference traffic. We cast prefix retrieval as aggregate-bandwidth scheduling across these tiers and present \proj, a KV cache engine that batches a step's reads, defers writes behind the forward pass, replicates each chunk on every tier, and lets each tier adaptively claim the next unread chunks once it is free. Our performance model shows this routing attains the aggregate-bandwidth bound without prediction or tuning, whereas any static split misses it by a margin that grows as per-tier bandwidths drift or fluctuate. It also predicts that the time-to-first-token (TTFT) gain grows with load, as faster retrieval drains the queue, and disappears when host memory or arrival order keeps reuse in cache. Plugged into LMCache and evaluated on four models, six datasets, and up to 128 GPUs, \proj retrieves within 5\% of the bound, delivers up to 1.49$\times$ higher offline throughput than LMCache, and lowers mean TTFT by up to 6.1$\times$ under bursty arrivals.
\end{abstract}

\keywords{Large Intermediate States of Transformer Models, Multi-path I/O Concurrency, Hierarchical Memory Management, I/O and Inference Performance Modelling.}

\maketitle

\section{Introduction}
\label{sec:intro}

Large language models (LLMs) now serve long-running agentic, coding, and document-analysis workloads that process tens to hundreds of thousands of tokens per request~\cite{llmapplication_liu2025comprehensive,contextlengthsummary_hsieh2024ruler,qwen3_yang2025qwen3,llama3_grattafiori2024llama}. The attention keys and values of every token, the KV cache, must be retained for all later tokens, so the KV cache grows with context length and dominates GPU memory. Requests frequently share long prefixes, such as system prompts, documents, code repositories, and the growing history of multi-turn and agentic sessions. The KV of prefixes are cached for later reuses, avoiding recomputations whose cost grows with model depth and quadratically with prefix length~\cite{attention_vaswani2017attention,attentioncompcomplexity_keles2023computational,mooncake_qin2024mooncake,cheng2025lmcache,nixon2026year}.

The set of unique prefixes that will be reused, the \emph{working set}, rarely fits on the GPU due to limited capacity. As new prefill and decode tokens need memory for their KV states, the inference engine evicts cached prefixes under policies such as LRU, LFU, or workflow-aware schemes~\cite{vllm_kwon2023efficient,sglang_zheng2024sglang,pan2025kvflow,ouyang2026unicache}. KV cache connectors such as LMCache and Mooncake~\cite{cheng2025lmcache,mooncake_qin2024mooncake} extend GPU capacity to slower but large memory and storage tiers, from which the prefix KVs are retrieved when later requests reuse them. Therefore, storage-backed prefix caching trades expensive GPU recomputation for I/O, and in this paper we characaterize, model, and accelerate the I/O bounds for faster inferencing.

\paragraph*{\textbf{Motivation for Storage-Tier Offloading of KV Cache}}
\label{sec:intro-motivation}

\begin{figure*}[!t]
    \minipage[t]{0.32\textwidth}
        \centering
        \includegraphics[width=\linewidth]{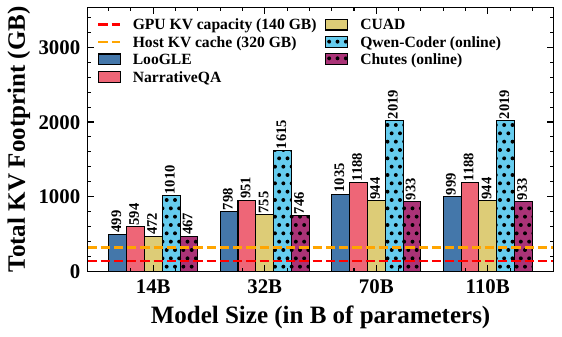}
        \captionsetup{skip=2pt,justification=raggedright}
        \caption{Working set (unique KV) vs.\ GPU and host KV budgets.}
        \label{fig:m1-kv-footprint}
    \endminipage
    \hfill
    \minipage[t]{0.32\textwidth}
        \centering
        \includegraphics[width=\linewidth]{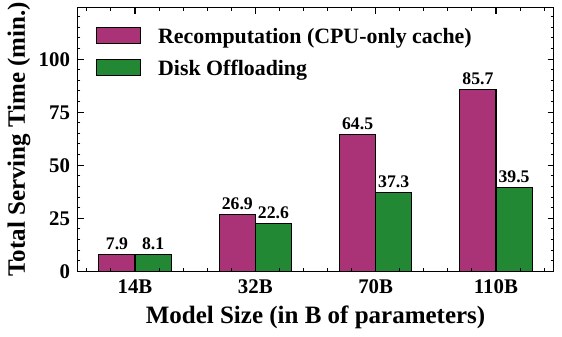}
        \captionsetup{skip=2pt,justification=raggedright}
        \caption{Serving time of LooGLE on \jlse: recompute vs.\ offload.}
        \label{fig:m2-offloading-wins}
    \endminipage
    \hfill
    \minipage[t]{0.32\textwidth}
        \centering
        \includegraphics[width=\linewidth]{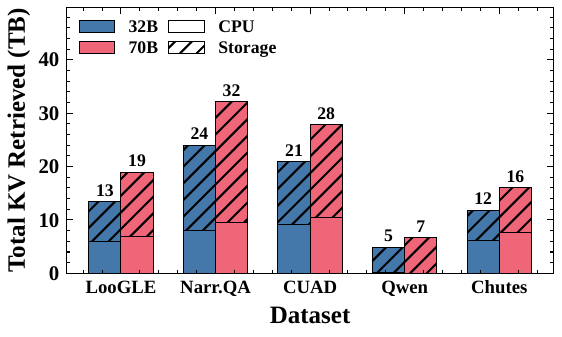}
        \captionsetup{skip=2pt,justification=raggedright}
        \caption{KV read per dataset from host memory and storage.}
        \label{fig:m4-kv-read-volume}
    \endminipage
\end{figure*}

Realistic long-context workloads have working sets (unique KV prefixes) far larger than GPU and host memory. On our testbeds (\S~\ref{sec:eval:setup}), four GPUs leave about 140~GB for KV after the model weights. Each node has 512~GB of host DRAM, 1.6--3.2$\times$ its GPU memory, similar to cloud GPU instances such as AWS \texttt{p4}, \texttt{p5}, and the new Nvidia-B300-based \texttt{p6}~\cite{AWS_P4,AWS_P5,AWS_P6}. Figure~\ref{fig:m1-kv-footprint} shows the working sets of three document datasets and two online production traces (detailed in \S~\ref{sec:eval:method}). Even though grouped-query attention (GQA)~\cite{gqa_ainslie2023gqa} already compacts the KV of every model, the working set of a 14B model (0.5--1~TB) is 3.3--7.2$\times$ the GPU KV budget, and at 110B it reaches $\sim$2~TB, up to 6.3$\times$ the host-memory KV cache of a node. We observe that the memory footprint of unique KV prefixes grows with model size due to more layers and attention heads. As a consequence, for the same dataset, the 14B model can hold the KV cache of larger number of prefix chunks than the 110B model.

We next see in Figure~\ref{fig:m2-offloading-wins}, for the LooGLE dataset, that despite disk reads being slow (6.2--13GB/s, Table~\ref{tab:testbeds}), utilizing LMCache~\cite{cheng2025lmcache} to store and fetch the offloaded prefixes is up to 2.2$\times$ faster then recomputing on H100 GPUs. This is because recomputation scales with model depth while retrieval scales only with the bytes moved. Furthermore, although each dataset's unique KV amounts to at most a few TBs (Figure~\ref{fig:m1-kv-footprint}), serving the entire dataset reads 5--32~TB for 32B and 70B models, as shown in Figure~\ref{fig:m4-kv-read-volume}. This read amplification is because eviction from the faster tiers (due to staleness) forces prefixes to be read again and again for reuse as the requests arrive in arbitrary order (from multiple users and concurrent sessions~\cite{nixon2026year}). Multiple efforts study and outline the benefit and operating envelope of storage offloading outperforming recomputations~\cite{kanichai2026pykvcache,hu2026bidaw,computeorstorage_jin2024compute}, and in this work, we focus on accelerating the storage regime to speedup prefix-reuse heavy workloads such as multi-turn sessions~\cite{narrativeqa_kocisky2018narrativeqa}, document analysis~\cite{cuad_hendrycks2021cuad}, coding agents~\cite{swebench_jimenez2024swebench}.

\paragraph*{\textbf{Limitations of the State of the Art}}
Existing systems take three directions to acclerate such prefix-reuse heavy workloads, and none systematically charaterizes and remove storage reads from the critical path once the working set (unique KV footprint) spills beyond the GPU and host memory (\S~\ref{sec:background:kv}). KV cache connectors such as LMCache, Mooncake, and NVIDIA Dynamo's KV block manager~\cite{cheng2025lmcache,mooncake_qin2024mooncake,nvidia_dynamo_kvbm_2026} extend the engine's cache with host memory and storage, but they read each matched chunk of each request in the batch in serial, non-overlapped fashion (detailed in \S~\ref{sec:gap-analysis}). Compression, token eviction, and selective restoration~\cite{cachegen_liu2024cachegen,h2o_zhang2023h2o,infinigen_lee2024infinigen} shrink each prefix by a constant factor, typically at the cost of accuracy. Modern GQA and MLA-based architecture already perform in-model KV compression, shrinking the KV sizes by 5--8$\times$ compared to standard multi-head attention for our evaluated models outlined in Table~\ref{tab:models}. Despite this with, as observed in Figure~\ref{fig:m1-kv-footprint}, the working set size (Unique KV prefix footprint) still spills beyond the GPU and CPU memory. Therefore, despite newer architectures, supporting large 1M+ context lengths (common for agentic workflows) for multiple concurrent users still requires memory capacity beyond what the GPU and CPU offer. Efforts such as Nvidia GDS~\cite{nvidia_gds}, Deepseek 3FS~\cite{an2024fireflyer}, IBM storage scale~\cite{ibm_redbooks_md260021_2026}, and Mooncake~\cite{mooncake_qin2024mooncake}, provide GPU-and/or filesystem-vendor specific optimizations, or proposing memory pooling capabilities for distributed KV caching, but do not design or model the I/O bounds of KV retrieval from multiple storage tiers on HPC systems. Lastly, KV reuse-aware schedulers such as HotPrefix~\cite{li2025hotprefix}, PEEK~\cite{xie2026peek}, Preble~\cite{srivatsa2025preble}, and Bidaw~\cite{hu2026bidaw} hide the storage retrieval overheads by shuffle the order of the requests to hide the storage retrieval overheads, which dishonors the arrival and/or priority order, and fails to accelerate if the prefix reuse order is not known apriori, which is typically the case for online serving.

\paragraph*{\textbf{Key Insight and Contributions}}
In this work, we leverage the insight the HPC and cloud datacenters are typically connected to fast remote storage such as a parallel file system~(PFS)~\cite{LustreFS} or AWS FSX~\cite{aws_fsx_lustre}, which can supplement additional storage bandwidth but remains idle during inferencing. To this end, we design \proj, which utilizes the I/O bandwidth of multiple storage tiers in parallel to accelerate prefix KV retrievals during inferences. We take a model-backed approach and cast prefix retrieval over heterogeneous storage tiers as an aggregate-bandwidth scheduling problem: within an iteration, the fastest that any engine can retrieve the prefixes it needs is set by the combined read bandwidth of the tiers it can read in parallel, the classic bound for dividing work among parallel servers~\cite{bharadwaj2003divisible}. \proj approaches this bound with \emph{adaptive multi-path completion-driven routing} and lands within 5\% of it. Our contributions are:

\begin{enumerate}[topsep=1pt,itemsep=1pt,leftmargin=12pt]
    \item \textbf{Characterization and gap analysis.} On two HPC testbeds, we show that storage-backed prefix retrieval is serialized, dominates step time, leaves the PFS bandwidth idle, and suffers read/write interference and heavy-tailed latency on the local and remote storage devices; and that host memory capacity, arrival order, and arrival rate decide whether it matters and how much (\S~\ref{sec:gap-analysis}).
    \item \textbf{Design and implementation of \proj.} We propose four mechanisms: step-level batched retrieval, deferred writes, multi-path replication, and completion-driven meet-in-the-middle routing, in a standalone C++ \texttt{io\_uring} engine that plugs into LMCache and transparently serves inference engines such as vLLM and SGLang (\S~\ref{sec:sys-design}).
    \item \textbf{Performance model of iteration breakdown and I/O bounds.} We decompose an iteration into retrieval, compute, and write; apply the classic aggregate-bandwidth bound of parallel servers to multi-tier retrieval; derive the penalty that a static split NVMe:PFS pays when the per-tier bandwidths deviate from it. We also explain the queueing delay and TTFT as functions of arrival order, arrival rate, host memory, and input and output lengths. We validate it with \emph{kvsim}, a trace-driven simulator that we release (\S~\ref{sec:model}).
    \item \textbf{Evaluation in offline and online serving.} With four models (14B--110B), six datasets, and two testbeds at up to 128 GPUs, \proj delivers up to 1.49$\times$ higher offline throughput than LMCache (3.22$\times$ over recomputation); for online real-world traces \proj lowers mean TTFT by up to 6.1$\times$ under bursty arrivals. We also characterize where \proj gains little or loses: its speedup over LMCache is 0.98--1.05$\times$ when host memory or the arrival order keeps reuse in cache, and 0.81$\times$ when reuse is sparse (\S~\ref{sec:evaluation},\S~\ref{sec:discussion}).

\end{enumerate}

\begin{figure*}
    \centering
    \includegraphics[width=\linewidth]{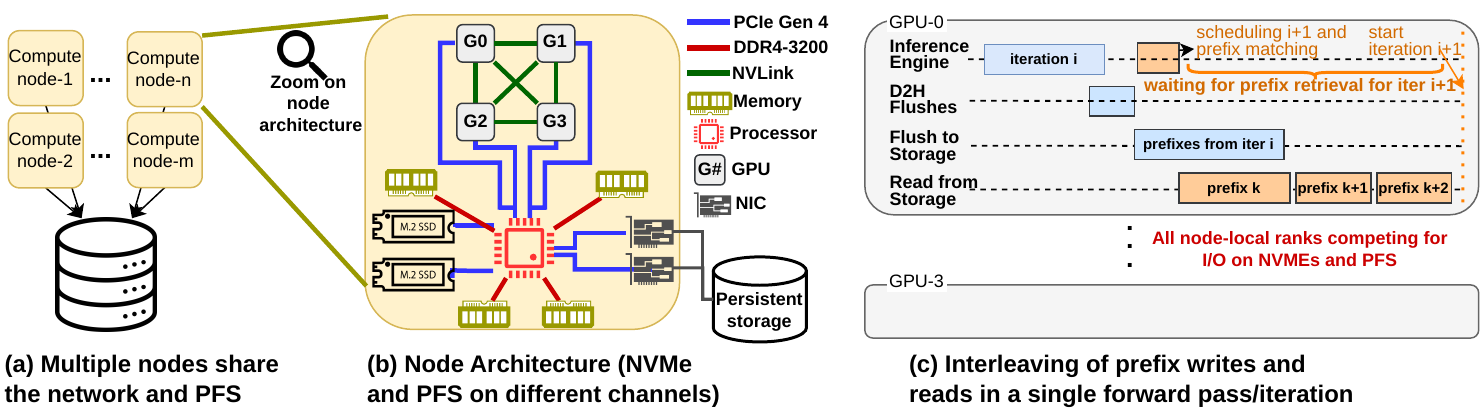}
    \caption{(a) Compute nodes share the bandwidth of a parallel file system (PFS). (b) A \polaris node: 4$\times$GPUs reach the NVMe and the PFS over independent paths. (c) Within one forward pass, each GPU's prefix writes for step $i$ interleave with its prefix reads for step $i{+}1$, and all ranks of a node compete for both tiers.}
    \label{fig:arch}
\end{figure*}

\paragraph*{\textbf{Scope and Limitations of Our Approach}}
\proj targets long-context workloads, such as agentic, coding, and document-analysis serving, whose unique prefix KVs spill beyond GPU and host memory and are reused heavily (81--96\% of prompt tokens hit a cached prefix in our datasets and traces). For workloads with small working sets, such as short conversations~\cite{burstgpt_wang2025burstgpt}, or with sparse reuse, storage leaves the critical path and the benefit of any storage-tier optimization~\cite{mooncake_qin2024mooncake,hu2026bidaw,li2025hotprefix,ouyang2026unicache,kanichai2026pykvcache,ren2025characterizing,computeorstorage_jin2024compute,nvidia_gds}, including ours, erodes (\S~\ref{sec:model:gains} states the conditions). Multi-path replication doubles the bytes written. Capacity is rarely the issue on multi-terabyte storage tiers: each tier independently evicts its chunks based on LRU when it nears its configured limit, and an evicted prefix is recomputed or fetched from alternate tier if it is requested again. The extra writes nevertheless cost coordination and bandwidth even when they are deferred off the read path, and when they cannot drain behind the forward pass they slow the iteration. If parallel reads do not recover more than replication costs, \proj is slower than single-tier offloading: 0.98$\times$ when host memory holds the working set (\S~\ref{sec:eval:order}), and 0.81$\times$ at 4\% reuse (\S~\ref{sec:eval:online}). \proj needs at least two storage tiers to realize its benefit; with one, it reduces to batched retrieval with deferred writes. \proj coordinates nothing across nodes: each node routes its own reads, so nodes that share the PFS adapt independently (\S~\ref{sec:eval:scaling}). Cross-node coordination, cross-node KV sharing, and prefill-decode disaggregation are orthogonal, and \S~\ref{sec:discussion} discusses how they compose with \proj.

\section{Background and Problem Formulation}
\label{sec:background}

\paragraph*{\textbf{KV caching across the memory hierarchy}}
\label{sec:background:kv}
Inference engines manage the KV cache in fixed-size blocks: vLLM's PagedAttention allocates it in pages to avoid fragmentation, and SGLang's RadixAttention and vLLM's automatic prefix caching index cached blocks by their token prefix, so that a new request reuses the longest cached prefix~\cite{vllm_kwon2023efficient,sglang_zheng2024sglang,APC}; CacheBlend extends reuse to cached chunks that are not prefixes~\cite{cacheblend}. FlexGen offloaded KV beyond the GPU for throughput-oriented inference~\cite{flexgen_sheng2023flexgen}. Today, connectors such as LMCache~\cite{cheng2025lmcache} extend the engine's prefix cache with host memory and storage, as CachedAttention does for multi-turn conversations~\cite{gao2024cachedattention}, Mooncake and NVIDIA Dynamo's KV block manager pool DRAM and SSDs across nodes for disaggregated prefill and decode~\cite{mooncake_qin2024mooncake,nvidia_dynamo_kvbm_2026}, and characterization studies of production traces and offloading stacks document both the reuse that makes it worthwhile to utilize storage tiers and the I/O that it generates~\cite{kvcache_wang2025kvcache,nixon2026year,ren2025characterizing}. On a prefix hit, these connectors read each matched chunk from the fastest tier that holds it and serve the requests of a step one after another.

\paragraph*{\textbf{Reducing and accelerating storage reads}}
Compression and quantization~\cite{cachegen_liu2024cachegen,shadowkv_sun2024shadowkv}, token eviction~\cite{h2o_zhang2023h2o,streamingllm_xiao2024efficient}, and selective or partial restoration~\cite{infinigen_lee2024infinigen,gao2025fast} shrink a prefix by a constant factor (3.5--4.3$\times$ for CacheGen), sometimes at a cost in accuracy. Architectures already build in much of this saving: GQA leaves 5--8$\times$ fewer KV heads in the models we evaluate, and multi-head latent attention cuts the KV cache of DeepSeek-V2 by 93.3\%~\cite{gqa_ainslie2023gqa,deepseekv2_2024}. Compression shrinks the spill but does not remove it, and it composes with our approach (\S~\ref{sec:discussion}). Storage backends make one tier faster: 3FS serves KV from RDMA-attached SSDs~\cite{an2024fireflyer}, GPUDirect Storage and NIXL move KV without host copies~\cite{nvidia_gds,nvidia_nixl}, py-kvcache and BiDAW drives NVMe with \texttt{io\_uring} and preloads the blocks of waiting requests~\cite{kanichai2026pykvcache,hu2026bidaw}, and IBM Storage Scale and Dynamo add shared storage as a further tier~\cite{ibm_redbooks_md260021_2026,nvidia_dynamo_kvbm_2026}. None reads a step's prefixes from several tiers at once. Reuse-aware schedulers, such as SGLang's longest-prefix-match~(LPM) policy, Preble, BiDAW, and PEEK, run requests that share a prefix while its KV is hot~\cite{sglang_zheng2024sglang,srivatsa2025preble,hu2026bidaw,xie2026peek}, and smarter eviction and promotion keep hot prefixes in the faster tiers~\cite{li2025hotprefix,ouyang2026unicache,pan2025kvflow}. Reordering may work for offline batches running without arrival or scheduling priorities, but online serving must respect arrival order, and better eviction helps only while the working set between reuses is not much larger than memory (detailed in \S~\ref{sec:model:demand}).

\begin{figure*}[ht]
    \minipage[t]{0.32\textwidth}
        \centering
        \includegraphics[width=\linewidth]{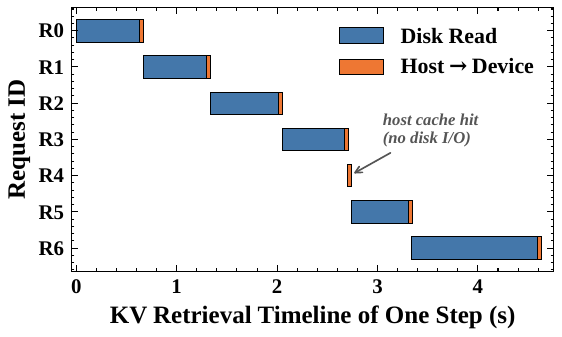}
        \captionsetup{skip=2pt,justification=raggedright}
        \caption{LMCache retrieval timeline of one step (70B): the transfers of each request run serially.}
        \label{fig:m5-retrieval-timeline}
    \endminipage
    \hfill
    \minipage[t]{0.32\textwidth}
        \centering
        \includegraphics[width=\linewidth]{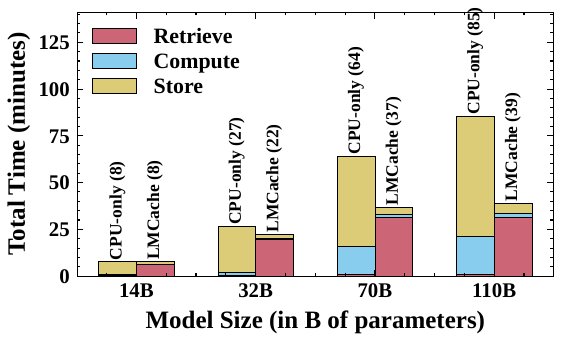}
        \captionsetup{skip=2pt,justification=raggedright}
        \caption{Processing-time breakdown of CPU-only and LMCache across model sizes (LooGLE, \jlse).}
        \label{fig:g02-baselines}
    \endminipage
    \hfill
    \minipage[t]{0.32\textwidth}
        \centering
        \includegraphics[width=\linewidth]{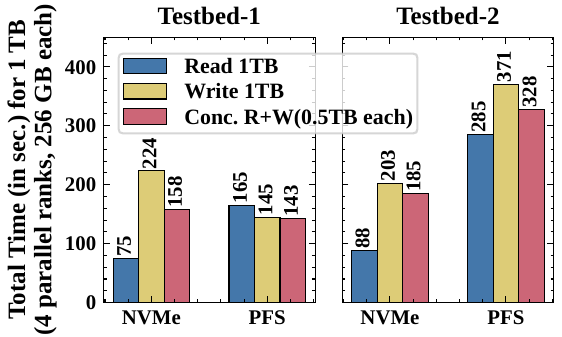}
        \captionsetup{skip=2pt,justification=raggedright}
        \caption{Mean per-rank time to read~(R) or write~(W) 1~TB alone, or 0.5~TB of each concurrently.}
        \label{fig:m3-io-char}
    \endminipage
\end{figure*}

\paragraph*{\textbf{The store and retrieve path}}
A connector such as LMCache splits each request's KV into chunks of 256 tokens by default and keeps one file per chunk and tensor-parallel rank on storage (Figure~\ref{fig:arch}(c)). Its \emph{store} path copies newly computed chunks from the GPU to a host buffer and flushes them to storage asynchronously, so that later requests sharing the prefix can reuse them. Its \emph{retrieve} path runs when the engine schedules the next step. The engine matches each request's prompt against the cached prefixes; matched chunks resident on the GPU need no transfer, chunks in host memory are copied to the GPU, and chunks only on storage are read into host memory and then copied. The forward pass cannot start on any tensor-parallel rank until all matched chunks of the batch are on its GPU, and the flushes of step $i$ overlap the reads of step $i{+}1$ on the same devices.

\paragraph*{\textbf{Node storage architecture}}
Figure~\ref{fig:arch}(a,b) shows the storage paths of a node of \polaris (Table~\ref{tab:testbeds}). Four GPUs, joined by NVLink, sit in two NUMA domains, each with its own PCIe links to host memory. From each domain, independent paths lead to the node-local NVMe drives (RAID-0) over PCIe and to the shared PFS through two NICs, so the two storage paths share no link before host memory. The layout also creates two kinds of contention: the tensor-parallel ranks of a job share the node-local NVMe, and the PFS is shared by all nodes and jobs of the cluster, so its bandwidth varies with unrelated load~\cite{yildiz2016rootcauses,VELOC-FGCS24}. Cloud GPU instances have the same shape, with instance-local NVMe next to network-attached file and object stores~\cite{AWS_P5,aws_fsx_lustre}.

\paragraph*{\textbf{Problem Formulation}}
\label{sec:formulation}

Inference proceeds in iterations (steps) indexed by $i$. Step~$i$ runs the forward pass of a batch of requests $B$, chosen by the engine's scheduler. Under tensor (or expert) parallelism, every participating GPU needs all matched prefix chunks of the batch before it can start, so the step is synchronous across ranks. We decompose its wall-clock time $T_i$ into three phases,
\begin{equation}
T_i \;=\; \Tret_i \;+\; \Tfwd_i \;+\; \Twr_i ,
\label{eq:step-decomp}
\end{equation}
where $\Tret_i$ retrieves the matched prefix chunks that are not already on the GPU, from host memory or storage; $\Tfwd_i$ executes the forward pass and depends only on the model and the batch; and $\Twr_i$ is the part of persisting newly generated KV that the step waits for, normally only the device-to-device copy, since flushes to host and storage run asynchronously. However, retrieval sits in the path of critical performance, causing slower $\Tret_i$ to directly transfer overheads on the iteration time $T_i$.

The two serving modes use this decomposition differently. Offline (batch) serving submits a whole dataset at once, and its throughput and end-to-end time are set by $\sum_i T_i$. Online serving admits requests as they arrive, and a request's time-to-first-token (TTFT) is the time it waits in the queue for previous batches/requests plus the amount of time taken by the step/batch to complete it's own prefill. A long step thus delays its own requests and every request queued behind it.

The problem is to minimize $\Tret_i$ at every step without exposing the overheads of asynchronous writes ($\Twr_i \approx 0$), under three constraints. First, the chunks that a step needs are known only when the scheduler forms the batch: they depend on which requests have arrived, in what order, and what the faster tiers still hold. Second, the read bandwidth of each storage device varies from step to step with concurrent writes, competing tensor-parallel ranks, and other jobs. Third, the engine's scheduling order is respected: a storage engine must accelerate the batches it is given rather than reorder requests to suit its cache (unlike BiDAW~\cite{hu2026bidaw}, PyKVCache~\cite{kanichai2026pykvcache}). \S~\ref{sec:gap-analysis} measures how existing systems fare on this problem, and \S~\ref{sec:model} returns to Equation~\ref{eq:step-decomp} to bound $\Tret_i$ based on our design.

\section{Gap Analysis and Observations}
\label{sec:gap-analysis}

We characterize storage-backed prefix retrieval in LMCache~\cite{cheng2025lmcache}, the state-of-the-art KV connector, on the testbeds and workloads of \S~\ref{sec:eval:setup}, offline (closed loop, throughput) and online (open-loop arrivals, TTFT). Observations O1--O4 concern how fast retrieval moves bytes; O5 and O6 concern whether storage retrieval matters for a given workload and deployment, and how much.

\paragraph*{\textbf{O1: Prefix retrieval is serialized request by request}}
LMCache retrieves a step's prefixes in a strictly sequential, per-request pipeline: for each request, all matched chunks are read from storage into host memory and copied to the GPU before the retrieval of the next request begins. Figure~\ref{fig:m5-retrieval-timeline} shows an Nsight Systems trace of one step that batches seven requests on a 70B model. The storage reads of each request take 600--1300~ms, the next request waits for them, and the step spends 4.7~s retrieving. NVMe controllers reach peak throughput only at high queue depths, when many I/O operations are in flight simultaneously~\cite{he2017unwritten,yan2017tinytail}. Issuing chunks one request at a time and alternating control between disk-to-host and host-to-device transfers, results in two limitations (a) the pipeline keeps the queue shallow and prevents the controller from exploiting its internal parallelism across channels and dies; and (b) it keeps hierarchical interconnects idle, i.e., when host-to-device link is used, the disk-to-host channel remains idle. Because the forward pass waits for the KV of every request, finishing one request early gains nothing; only the arrival of the step's last chunk matters.

\begin{figure*}[!ht]
    \minipage[t]{0.32\textwidth}
        \centering
        \includegraphics[width=\linewidth]{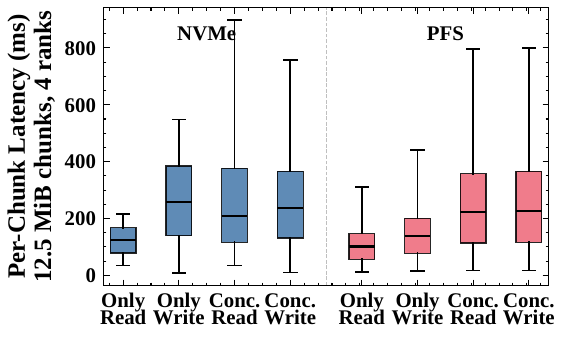}
        \captionsetup{skip=2pt,justification=raggedright}
        \caption{Per-chunk I/O latency of reads and writes alone (`Only') and concurrent (`Conc.', 50:50).}
        \label{fig:m6-io-latency}
    \endminipage
    \hfill
    \minipage[t]{0.32\textwidth}
        \centering
        \includegraphics[width=\linewidth]{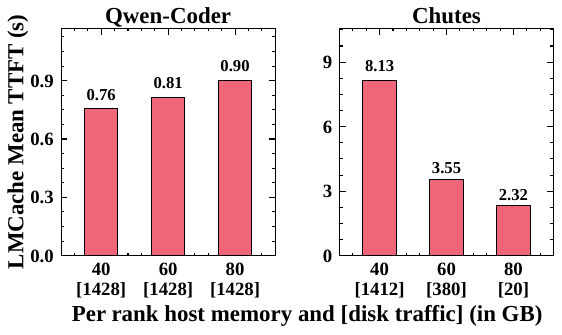}
        \captionsetup{skip=2pt,justification=raggedright}
        \caption{LMCache mean TTFT against host memory; brackets give storage reads per rank.}
        \label{fig:o5-memory}
    \endminipage
    \hfill
    \minipage[t]{0.32\textwidth}
        \centering
        \includegraphics[width=\linewidth]{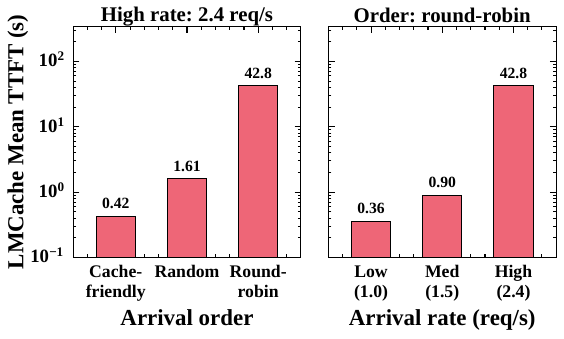}
        \captionsetup{skip=2pt,justification=raggedright}
        \caption{LMCache mean TTFT on Qwen-Coder against arrival order (left) and arrival rate (right).}
        \label{fig:o6-order-rate}
    \endminipage
\end{figure*}

\paragraph*{\textbf{O2: Prefix retrieval dominates step time}}
Figure~\ref{fig:g02-baselines} decomposes the processing time of the CPU-only baseline, which keeps KV only in GPU and host memory and recomputes evicted prefixes, and of LMCache, which offloads them to NVMe, for four model sizes, each with a different KV footprint per token (Table~\ref{tab:models}). Because of the read amplification of Figure~\ref{fig:m4-kv-read-volume}, retrieval dominates LMCache's time at every size, accounting for 80--88\% of it. Even so, LMCache matches CPU-only at 14B and finishes up to 2.2$\times$ sooner at larger models, where CPU-only, which must recompute every evicted prefix, spends far longer in its again computing and repeatedly storing it. Since retrieval is the major fraction of the total serving time, accelerating it speeds up inferences.

\paragraph*{\textbf{O3: Second-tier bandwidth sits idle while the local device suffers read/write interference}}
To measure the bandwidth available to KV offloading, we run an \texttt{io\_uring} MPI microbenchmark that mirrors the inference pattern: four ranks, matching the tensor-parallel degree, issue \texttt{O\_DIRECT} chunk reads and writes, one file per chunk, in isolation and concurrently (Figure~\ref{fig:m3-io-char}). Two observations follow. First, in isolation the node-local NVMe reads only 2.2$\times$ (\jlse) and 3.2$\times$ (\polaris) faster than the PFS, which is shared by all compute nodes. Yet during inference the PFS carries no traffic beyond model loading, so its bandwidth is available for free. On cloud providers, fast storage tiers such as AWS FSx~\cite{aws_fsx_lustre} support tens of GB/s of throughput cheaper than GPUs, making inferences both faster and cheaper than using more GPUs for recomputations. Second, when one device serves reads and writes together, as interleaved retrieval and flushing force it to, the two interfere: 1~TB of concurrent reads and writes (512~GB each) takes 6\% longer on \jlse and 27\% longer on \polaris than running them in isolation. The read bandwidth a device delivers during inference therefore depends on what else it is doing, and a division of reads across devices computed from their isolated bandwidths is mis-sized for the concurrency in which inference runs.

\paragraph*{\textbf{O4: Per-chunk latency is heavy-tailed and per-tier bandwidth fluctuates}}
Even a well-chosen division of reads meets unstable per-chunk latency, especially on the shared PFS. On \polaris with four ranks (Figure~\ref{fig:m6-io-latency}), isolated NVMe reads have a p50 of 123~ms and a p99 of 216~ms; under concurrent reads and writes the p99 reaches 898~ms, 7.3$\times$ the isolated median, and PFS reads show a p99 of 795~ms against a 102~ms median. The tails have different origins: write amplification and garbage collection on the NVMe~\cite{he2017unwritten,yan2017tinytail}, and cluster-wide contention on the PFS (Figure~\ref{fig:arch}(a))~\cite{yildiz2016rootcauses,VELOC-FGCS24}. Together with the tensor-parallel ranks that contend for reads and writes on the NVMe, they make the chunk retrieval latency and bandwidth fluctuate across different inference steps, making static split of prefixes across storage tiers suboptimal (detailed in \S~\ref{sec:model}).

\paragraph*{\textbf{O5: Host memory relative to the working set decides whether storage is on the critical path}}
Whether storage reads happen at all depends on how much KV is touched between two uses of a prefix compared with how much host memory holds. Figure~\ref{fig:o5-memory} varies host memory per rank while the 32B model serves two online production traces (\S~\ref{sec:eval:method}): Qwen-Coder and Chutes at their medium arrival rates (50\% of times vLLM's scheduling queue was idle, and requests didn't have to wait, detailed in \S~\ref{sec:eval:method}). On Qwen-Coder, with sessions interleaved round-robin, the KV touched between reuses far exceeds host memory: doubling it from 40 to 80~GB per rank changes the storage read volume by less than 1\% (1428.3 to 1428.2~GB per rank), and LMCache's mean TTFT stays between 0.76 and 0.90~s. On Chutes, the same doubling brings almost every reuse within memory: storage reads fall by 98.6\% (1412 to 20~GB per rank), and LMCache's mean TTFT falls 3.5$\times$, from 8.13 to 2.32~s. This analysis demonstrates two sub-observations: (a) when the working set between reuses is large (beyond host memory capacity), and only faster storage retrieval can; and (b) When it fits in the faster tiers, storage leaves the critical path, and no storage-side optimization (e.g., BiDAW~\cite{hu2026bidaw}, 3FS~\cite{an2024fireflyer}, PyKVCache~\cite{kanichai2026pykvcache}, etc.), ours included, can accelerate the workload. Appendix~\ref{sec:appendix:demand} adds \proj to this sweep (Figure~\ref{fig:o8-dram}).

\paragraph*{\textbf{O6: Arrival/Scheduling order and arrival rate shape both the demand and the queue}}
Figure~\ref{fig:o6-order-rate} (left) serves the same Qwen-Coder trace with the 32B model at a high arrival rate (2.4~req/s, \S~\ref{sec:eval:rates}), with the same output lengths and host memory (80~GB per rank), in three arrival orders. In the cache-friendly order of the source trace (similar to BiDAW~\cite{hu2026bidaw}, SGLang's LPM, PEEK~\cite{xie2026peek}, and PyKVCache~\cite{kanichai2026pykvcache}), consecutive turns of a session arrive close together: the median reuse distance, the KV touched between two uses of a chunk, is 13~GB per rank, far below host memory (80~GB/rank), because of which only 291.3~GB (instead of 1692~GB) per rank is read from storage, leading to LMCache's mean TTFT of 0.42~s. Interleaving whole sessions at random (representative of real-world multi-session concurrent serving~\cite{nixon2026year}), with turn/request order per-session preserved, raises the median reuse distance to 67~GB and storage reads to 1300~GB per rank, leading to mean TTFT of 1.61~s. And finaly, interleaving them round-robin (giving each concurrent session fair/equal priority) raises it to 278~GB and 1692~GB per rank, leading to 42.8~s mean TTFT. Order is a property of the arrivals, and the more sessions it interleaves between two turns of one session, the longer every reuse distance. Next, Figure~\ref{fig:o6-order-rate} (right) fixes the order (round-robin) and raises the arrival rate from low through medium to high. The mean number of requests in flight grows from 3.4 to 128, and LMCache's mean TTFT grows 120$\times$, from 0.36 to 42.8~s, most of it waiting behind slow retrievals. Scheduling or arrival order sets the volume to be read from storage, and rate sets how many requests wait behind each read. A server that must respect arrival order cannot choose the cache-friendly order; it can only make the reads it is given faster (extended in Appendix~\ref{sec:appendix:demand}).

\paragraph*{\textbf{Takeaways}}
O1--O4 expose losses that a storage engine can recover: retrieval is serialized per request (O1) and dominates the step (O2), the PFS bandwidth sits idle while the local device suffers read/write interference (O3), and per-tier bandwidth fluctuates with heavy tails (O4). O5 and O6 lie outside any storage engine's control: host memory, arrival/scheduling order, and arrival rate decide how much of the working set reaches storage and how many requests wait behind each read, and therefore how much a faster storage read path can speed up the inference.

\section{\proj System Design and Implementation}
\label{sec:sys-design}

\proj minimizes the retrieval term $\Tret_i$ that dominates each step (O2) under three design goals: it serves batches in the order the engine schedules them, without reordering requests to suit its cache; it predicts nothing about future demand or bandwidth; and it needs no tuning at run time. Four I/O mechanisms address the gaps of \S~\ref{sec:gap-analysis}: M1 pools a step's storage reads (O1), M2 keeps writes off the devices while a step reads (O3), M3 replicates every chunk on every tier to use the idle PFS (O2, O3), and M4 routes each read by completion rather than by a fixed split (O4).

\subsection{\textbf{\proj's I/O Mechanisms}}
\label{sec:design}

\paragraph{\textbf{M1: Parallel, batched prefix retrieval}}
\label{sec:design:joint-scheduling}
A forward pass batches many requests to keep the GPU busy and cannot begin until the prefixes of all of them are on the GPU. \proj therefore treats a step's storage reads as one pool rather than as a sequence of per-request pipelines (O1). It classifies the matched chunks of all requests by where they reside, submits every storage-resident chunk into one parallel I/O queue, and drains it with a single completion loop. Pooling is what makes the remaining mechanisms possible: the pool can be divided among tiers, and a tier that finishes early can take more of it. Pooling also simplifies I/O management in three ways. Buffers are allocated in bulk, removing the per-request block-management and scheduling overhead that a one-at-a-time pipeline pays repeatedly. One completion-drain loop serves all requests, so reads finish in any order without per-request synchronization. And host-to-device transfers pipeline naturally: each chunk ships to the GPU as soon as it arrives, overlapping the PCIe transfer with the remaining storage reads instead of alternating between the two.

\paragraph{\textbf{M2: Deferred writes}}
\label{sec:design:deferred-writes}
The KV generated in a step must be persisted for future reuse. Under tensor parallelism, all ranks finish together and flush their KV at once, competing for write bandwidth. Meanwhile, the next step needs a different set of prefixes read back, and those reads are on the critical path. Letting the writes run against these reads delays retrieval, and read/write interference amplifies the delay (O3, Figure~\ref{fig:arch}(c)). \proj prioritizes reads and defers writes: new KV is buffered in host memory and flushed during the following forward pass, when the devices are otherwise idle for reads. A retrieval is always followed by a forward pass that leaves such an idle window, and the window is usually long enough for a step's writes to drain. Deferral moves writes in time but does not remove them: every chunk is still written, and a flush that outlasts the forward pass overlaps the next step's reads. \S~\ref{sec:model:iter} states when this happens, and \S~\ref{sec:eval:envelope} measures how often.

\paragraph{\textbf{M3: Multi-path replication}}
\label{sec:design:replication}
When requests arrive in an order the server does not control, which chunks a step will need cannot be predicted, so any placement that commits each chunk to one tier risks loading too much of a step's demand onto one tier and too little onto another, and the bandwidth each tier will deliver at read time is unknown as well (O4). Rather than guess, \proj writes each chunk to every tier. Storage capacity is plentiful relative to the working set, and because writes are deferred behind the forward pass, the second copy normally stays off the read path; it still consumes write bandwidth, which \S~\ref{sec:model:gains} weighs against the read bandwidth it unlocks. Replication gives the engine full flexibility: any part of any prefix can be read from whichever tier is free, which enables the routing below and lets a step draw on the aggregate, otherwise idle bandwidth of all tiers (O3). The scheme is not specific to NVMe and PFS; any further tier, such as CXL-attached memory or a second NVMe, adds read bandwidth with no placement or migration policy to maintain.

\paragraph{\textbf{M4: Completion-driven meet-in-the-middle routing}}
\label{sec:design:work-stealing}
Replication lets \proj read any part of any prefix from any tier, which it exploits with a meet-in-the-middle strategy. For two tiers, it reads the beginning of a prefix from the NVMe and the end from the PFS, in batches of $m$ chunks, and lets the two directions meet in the middle, as Cake does between loading and recomputing KV~\cite{computeorstorage_jin2024compute}. A tier that finishes its batch immediately pulls the next outstanding chunks, so a faster or momentarily less contended tier advances its cursor further and takes on more work. This has four consequences: each tier stays saturated, since neither waits on the other; the scheduling and polling overhead of reaping many small one-by-one reads is avoided; tail latency is masked, because both tiers rarely stall at once, so the healthy tier absorbs the stragglers of O4; and no bandwidth measurement, prediction, or interleaving schedule is needed, since the engine adapts purely by completion. The cursors need not synchronize at the meeting point: a small overlap re-reads a few chunks into the same host buffer, and because the content is identical, the first arrival makes the chunk GPU-ready while the redundant read completes harmlessly. \S~\ref{sec:model:bound} shows that this routing attains the aggregate-bandwidth bound.

\subsection{\textbf{I/O Lifecycle of an Iteration}}
\label{sec:algorithm}

\begin{algorithm}
    \small
    \caption{I/O lifecycle of one iteration}
    \label{algo:kvstream-lifecycle}
    \DontPrintSemicolon
    \SetKwInOut{KwIn}{Input}
    \SetKwFunction{EnqueueWrites}{enqueue\_writes}
    \SetKwFunction{FlushWrites}{flush\_writes}
    \SetKwFunction{PrefixMatch}{prefix\_match}
    \SetKwFunction{SubmitReads}{submit\_reads}
    \SetKwFunction{InitiateHtoD}{h2d}
    \SetKwFunction{WaitAny}{wait\_any}
    \SetKwFunction{StealNext}{claim\_next}
    \SetKwFunction{ForwardPass}{forward}
    \KwIn{batch $B$ of iteration $i{+}1$; writes $\mathcal{W}_i$ deferred from iteration $i$; batch size $m$}
    \EnqueueWrites{$\mathcal{W}_i$} \tcp*{stage on all tiers; don't submit}
    $C_\mathrm{gpu}, C_\mathrm{cpu}, C_\mathrm{disk} \gets$ \PrefixMatch{$B$} \tcp*{all reqs. at once}
    \SubmitReads{$C_\mathrm{disk}$, $m$} \tcp*{NVMe from front, PFS from back}
    \lForEach{$c \in C_\mathrm{cpu}$}{\InitiateHtoD{$c$}}
    \While{\textnormal{not all chunks are on the GPU}}{
        $c, \mathit{src} \gets$ \WaitAny{}\;
        \If{$\mathit{src}$ \textnormal{is a storage tier}}{
            \InitiateHtoD{$c$}\;
            \StealNext{$\mathit{src}$, $m$} \tcp*{next $m$ unread chunks}
        }
        mark $c$ GPU-ready\;
    }
    \FlushWrites{} \tcp*{submit the staged writes}
    \ForwardPass{$B$} \tcp*{overlaps the writes}
\end{algorithm}
Algorithm~\ref{algo:kvstream-lifecycle} combines the four mechanisms in one iteration. It runs on each tensor-parallel rank, and the inference engine drives the iterations. Line~1 stages the writes deferred from the previous iteration on every tier (\S~\ref{sec:design:replication}) but does not submit them, so the reads that follow meet no write contention. Line~2 matches the prefixes of all requests in the batch that the engine scheduled~\cite{KVCacheProfile-IPDPS25} and classifies each matched chunk by where it resides: on the GPU, in host memory, or only on storage. Line~3 submits all storage-resident chunks as one pool. The NVMe reads from the front of the list and the PFS from the back, $m$ chunks at a time, while Line~4 starts copying host-resident chunks to the GPU. Lines~5--10 drain completions from any source. When a storage read completes, its chunk moves to the GPU, and its tier claims the next $m$ unread chunks (Line~9). Near the meeting point, the tiers read at most $2m$ chunks twice. Line~11 submits the staged writes. The forward pass (Line~12) overlaps them and the device-to-host copies of the KV it produces, which the next iteration stages in Line~1.

\subsection{\textbf{Implementation}}
\label{sec:impl}
\proj's I/O engine is a standalone C++ library with no CUDA dependency, so it carries over to other GPU architectures. It issues asynchronous \texttt{O\_DIRECT} reads and writes through Linux's \texttt{io\_uring}, with separate rings for reads and writes, so that waiting on one never stalls the other. A \texttt{nanobind} binding exposes it to Python with zero-copy buffers. \proj plugs into LMCache~\cite{cheng2025lmcache} as a storage backend: the adapter between LMCache and the inference engine calls \proj's batch-level API at the start of each iteration, during prefix matching. The engine keeps using LMCache's existing interface, so vLLM~\cite{vllm_kwon2023efficient} and other engines that integrate LMCache get multi-tier retrieval, completion-driven routing, and deferred writes without code changes. In SGLang~\cite{sglang_zheng2024sglang}, \proj is a storage backend of the hierarchical cache (HiCache). There it writes synchronously and does not overlap retrieval, because HiCache reuses a host page as soon as a write returns and finishes or abandons a prefetch before it schedules the request.

\section{Performance Model of Iteration Breakdown and I/O Bounds}
\label{sec:model}
\label{sec:design:model}

We now return to the iteration of Equation~\ref{eq:step-decomp} to answer three questions: how many bytes a step must read from storage (\S~\ref{sec:model:demand}), how fast several tiers can read them and why a static split cannot (\S~\ref{sec:model:bound}), and when faster reads improve end-to-end performance (\S~\ref{sec:model:gains}). 
Table~\ref{tab:model-map} maps each workload and system factor to the term of the model that it moves, the observation that motivates it, and the experiment that tests it. 

\begin{table}[t]
\centering
\footnotesize
\setlength{\tabcolsep}{4pt}
\caption{Factors of the performance model: the term each moves, where it is observed, and evaluated.}
\label{tab:model-map}
\begin{tabularx}{\linewidth}{X|X|c|c}
\hline
\textbf{Workload or system factor} & \textbf{Model term} & \textbf{Observed} & \textbf{Evaluated} \\
\hline
Host memory capacity & demand $D$ via capacity $M$ (\S~\ref{sec:model:demand}) & O5 & \S~\ref{sec:eval:envelope} \\
Arrival order & demand $D$ via reuse distance $d$ (\S~\ref{sec:model:demand}) & O6 & \S~\ref{sec:eval:envelope} \\
Input length, prefix reuse, KV per token & demand $D$ via reused volume $V$ (\S~\ref{sec:model:demand}) & O2 & \S~\ref{sec:eval:offline} \\
Tier bandwidth, R/W and TP contention, tails & bound and static penalty (\S~\ref{sec:model:bound}) & O3, O4 & \S~\ref{sec:eval:validation} \\
Output length & forward time $\Tfwd$, retrieval share (\S~\ref{sec:model:gains}) & O2 & \S~\ref{sec:eval:envelope} \\
Arrival rate and burstiness & queueing (\S~\ref{sec:model:gains}) & O6 & \S~\ref{sec:eval:online} \\
Replicated writes & write path $\Twr$, write cost (\S~\ref{sec:model:gains}) & O3 & \S~\ref{sec:eval:envelope} \\
\hline
\end{tabularx}
\end{table}

\subsection{\textbf{Retrieval Demand as Function of Reuse Distance and Memory Capacity}}
\label{sec:model:demand}

Let the retrieval demand $D_i$ of step~$i$ be the bytes of its matched prefixes, per rank, that miss both GPU and host memory and must be read from storage. Over a serving session, $D = \sum_i D_i$ is set by two properties of the workload and one of the deployment. The workload reuses a volume $V$ of prefix KV in total, the matched prefix tokens times the per-rank KV bytes per token, so $V$ grows with input length, prefix reuse, and model size (Table~\ref{tab:models}). Whether a reused chunk is still in memory depends on its \emph{reuse distance} $d$: the bytes of distinct KV touched between two consecutive uses of the chunk, that is, the working set between those uses. Under LRU eviction a reuse hits in memory exactly when $d$ does not exceed the GPU+host cache capacity $M$, so
\begin{equation}
D \;=\; V \cdot \Pr[\, d > M \,],
\label{eq:demand}
\end{equation}
where the probability ($\Pr$) is taken over all reuses in the session. Other policies, such as LFU or workflow-aware eviction~\cite{pan2025kvflow,ouyang2026unicache}, change which chunks miss but not the dependence of $d$ on $M$.

Equation~\ref{eq:demand} separates what the deployment controls from what it does not. $M$ is a provisioning choice; $d$ is set by the dataset and by the order in which requests arrive or get scheduled, and interleaving more sessions between two turns of the same session lengthens reuse distance. One rule thus explains both O5 and O6 (\S~\ref{sec:gap-analysis}). When nearly all reuse distances exceed $M$, as on Qwen-Coder (Figure~\ref{fig:o5-memory}), where $d$=278~GB per rank, against $M=$40--80~GB, we get $\Pr[d > M] \approx 1$ at every memory size, and only faster retrieval helps. However, for the Chutes dataset (Figure~\ref{fig:o5-memory}), where $d$=58GB, increasing $M$ from 40~GB to 80~GB, reduces most of the storage demand.
We evaluate Equation~\ref{eq:demand} against measured reads in \S~\ref{sec:eval:workingset}.

\subsection{\textbf{Aggregate-Bandwidth Bound and the Static-Split Penalty}}
\label{sec:model:bound}
\label{sec:model:static}

Retrieving $D_i$ bytes requires fetching from storage-to-host and then host-to-device, whose bandwidth we denote by $\bbus$. Let $b_k(i)$ be the read bandwidth that tier~$k$ delivers to one rank during step~$i$, under that step's concurrent writes and tensor-parallel contention (not the bandwidth of an idle tier). Because the NVMe and PFS paths are independent (\S~\ref{sec:background:kv}, Figure~\ref{fig:arch}), their read bandwidths are additive, and no retrieval schedule $\Tret_i$ can read faster than $\underline{T}_i$, the \emph{aggregate-bandwidth bound}, which is the classic optimum for dividing work among parallel servers, which finish together when each takes a share proportional to its current speed~\cite{bharadwaj2003divisible,patterson1988raid,rodriguez2002dynamic}:

\begin{equation}
\Tret_i \;\ge\; \underline{T}_i \;=\; \frac{D_i}{\min\!\big(b_0(i)+b_1(i),\; \bbus\big)} .
\label{eq:bound}
\end{equation}

\proj applies it within every inference step, where the per-tier bandwidths change from step to step. Since device to host bandwidth $\bbus$ is typically magnitudes of order faster than storage tiers, the \emph{effective aggregate-bandwidth bound} becomes $\underline{T}_i = D_i/(b_0(i)+b_1(i))$. We measure a given retrieval policy (e.g., static 80:20 from NVMe:PFS) by its \emph{gap}, $\sum_i \Tret_i / \sum_i \underline{T}_i - 1$, both of which are computed based on bandwidth achieved at step $i$. 

A policy reaches the bound exactly when it is \emph{work-conserving} across tiers, i.e., no tier is idle until all chunks for that step $i$ are retrieved. \proj's routing is work-conserving by construction. Every chunk is on both tiers, and a tier that completes a batch immediately claims the next unread chunks from its end of the step's list (Line~9 of Algorithm~\ref{algo:kvstream-lifecycle}), so a momentarily faster tier reads more and the two tiers finish together. At step $i$, the share $\varphi$ of the demand $D$ that gets served by the faster tier (assuming tier-0 is fast, i.e., $b_0(i)> b_1(i)$) is given by
\begin{equation}
\varphi(i) \;=\; \frac{b_0(i)}{b_0(i)+b_1(i)},
\label{eq:phi}
\end{equation}
without any monitoring of bandwidths to redistribute the reads, and the retrieval time of \proj's completion-driven routing is therefore $\underline{T}_i \;=\; D_i/(b_0(i)+b_1(i))$, execept the only residual is each tier's last batch of at most $m$ meet-in-the-middle chunks (\S~\ref{sec:design:work-stealing}, M4).

A static policy commits each chunk to one tier when it is written, so the fraction $g$ of step~$i$'s demand that the fast tier serves is fixed by placement rather than by the bandwidths at read time. The tiers read in parallel, cannot help the other finish, so the step ends when the slower one does,
\begin{equation}
T^{\mathrm{stat}}_i \;=\; \max\!\Big(\frac{g\,D_i}{b_0(i)},\; \frac{(1-g)\,D_i}{b_1(i)}\Big),
\label{eq:static}
\end{equation}
and dividing by the aggregate-bandwidth bound gives its penalty,
\begin{equation}
\frac{T^{\mathrm{stat}}_i}{\underline{T}_i} \;=\; \max\!\Big(\frac{g}{\varphi(i)},\; \frac{1-g}{1-\varphi(i)}\Big) \;\ge\; 1,
\label{eq:penalty}
\end{equation}
with equality only when $g = \varphi(i)$ for step $i$. 
Each argument of Equation~\ref{eq:penalty} is a tier's load relative to its share of the bandwidth, so a split tuned to one bandwidth mix pays when the mix moves in either direction. A 75:25 split meets the aggregate-bandwidth bound for step $i$ if NVMe and PFS have 3~GB/s and 1~GB/s read throughputs, respectively; but it is 12.5\% slower than the bound when the NVMe slows to 2~GB/s, and slower than the NVMe alone when the PFS slows to 0.5~GB/s. 

The penalty is first-order in the deviation, $|\varphi-g|/g$ when the fast tier lags and $|\varphi-g|/(1-g)$ when it leads, so the session-average penalty grows roughly linearly with the mean deviation of $\varphi(i)$ from $g$. The share deviates for three reasons (O3, O4): tensor-parallel ranks contend for the NVMe, which also absorbs writes; other jobs and heavy tails modulate the PFS; and bursty arrivals land bursts of writes on the devices while they read. \proj's completion-driven routing removes the coupling using bandwidth centric instantaneous read distribution $\varphi(i)$, so Equation~\ref{eq:penalty}'s $T^{\mathrm{stat}}_i\approx\underline{T}_i$ for varying bandwidths. We evaluate in \S~\ref{sec:eval:validation} both predictions: the gap of a static split widens with the deviation of $\varphi(i)$ from $g$, while completion-driven routing stays at the bound.

\subsection{\textbf{End-to-End Gains and Losses}}
\label{sec:model:gains}
\label{sec:design:assembly}
\label{sec:model:iter}

Equations~\ref{eq:bound}--\ref{eq:penalty} concern a single step. However, in real-world deployments serving continuous stream of requests, the effective overall savings or overheads achieved by \proj (or other storage-tier optimizations) depend on the following.

\paragraph*{\textbf{Service time}} Routing a step's reads over both tiers instead single tier reduces retrieval from $D_i/b_0(i)$ to $D_i/(b_0(i)+b_1(i))$, that is, by a fraction $1-\varphi(i)$ of the single-tier's retrieval time. However, since a step's time $T_i$ is composed of both $\Tret$ and the time for forward pass $\Tfwd$ (Equation~\ref{eq:step-decomp}), the gain of step $i$ is capped by retrieval's share of the step (Amdahl's law~\cite{amdahl1967validity}): large when retrieval dominates, as in O2, and negligible when most matched prefixes come from memory or when long outputs or decodes stretch $\Tfwd$. In offline serving, the engine is saturated and throughput is inversely proportional to step time, so a closed-loop run measures this capacity gain.

\paragraph*{\textbf{Queueing}} In online serving, a request's TTFT is its own step time plus the time it waits in the scheduling queue for previous requests to complete. 
Faster retrieval therefore helps twice: the batch that retrieves finishes sooner, and the queue behind it drains faster, so every waiting request starts sooner. At low load, the TTFT gain is the per-request saving due to faster retrievals, which is small; however, at moderate or higher loads, when baseline approaches experience saturation due to slower storage reads, faster draining of waiting queue with \proj accumulates the TTFT gains across all batches, accelerating real-time serving. 

\paragraph*{\textbf{Losses}} The cost of replication does not shrink with demand. Let $W_i$ be the volume of new KV that step~$i$ persists per rank (fixed by workload) and $w_k$ the per-rank write bandwidth of tier~$k$. 
The newly generated unique KV cache gets flushed from the GPU using asynchronous transfers between GPU, host, and storage tier (\S~\ref{sec:design:deferred-writes}). The writes stay off the critical path as long as $W_i/\min_k w_k \lesssim \Tfwd_{i+1}$ on average, with the buffer absorbing short bursts. This happens when arrivals leave idle gaps and when steps append little new KV to long reused prefixes (multi-turn conversations~\cite{mooncake_qin2024mooncake} and agentic workflows~\cite{swebench_jimenez2024swebench,kvcache_wang2025kvcache}). Otherwise writes impact critical path in two ways: \emph{interference}, when reads and writes share the devices and each tier delivers less read bandwidth (O3), and \emph{backpressure}, when the staging buffer fills and the device-to-host copy of new KV waits, which lengthens $\Twr_i$. A second tier therefore pays when storage reads are large relative to the replicated writes, and it gains little or loses when they are small: when the working set between reuses fits in memory (O5), when the arrival order keeps reuse distances short (O6), or when reuse is sparse. \S~\ref{sec:eval:order} measures the gains and losses corrresponding to these effects.

\section{Performance Evaluation}
\label{sec:evaluation}
\label{sec:eval}

\subsection{\textbf{Experimental Setup}}
\label{sec:eval:setup}
\label{sec:eval:approaches}
\label{sec:eval:method}

\noindent\paragraph{\textbf{Testbeds}} Table~\ref{tab:testbeds} summarizes the two HPC testbeds. Their host-to-GPU memory ratios, 1.6:1 and 3.2:1, are similar to AWS \texttt{p6}, \texttt{p5}, and \texttt{p4} instances~\cite{AWS_P6,AWS_P4,AWS_P5}. Both testbeds have 2$\times$1.6~TB NVMe SSDs in RAID-0 as node-local storage, and every approach uses this array as its NVMe tier, so single-tier offloading already stripes its reads over two drives. \jlse and \polaris mount a 4~PB VAST PFS and 100~PB Lustre file systems (HPE ClusterStor E1000), respectively. The software stack is vLLM~\cite{vllm_kwon2023efficient} v0.15.0, LMCache~\cite{cheng2025lmcache} v0.3.14, and PyTorch v2.9.1 with CUDA v12.8.

\begin{table}[t]
\centering
\footnotesize
\setlength{\tabcolsep}{3pt}
\caption{Testbed configurations used in evaluations (R/W: read/write bandwidth).}
\label{tab:testbeds}
\begin{tabular}{l|c|c}
\hline
\textbf{Feature} & \textbf{\jlse} & \textbf{\polaris} \\
\hline
GPUs per node & 4$\times$H100-80GB & 4$\times$A100-40GB \\
Host memory (GB) & 512 & 512 \\
D2H/H2D (GB/s) & 55 & 25 \\
\hline
NVMe (RAID-0) & 2$\times$1.6~TB & 2$\times$1.6~TB \\
NVMe R/W (GB/s) & 13.7 / 4.6 & 11.6 / 5.1 \\
\hline
Parallel FS & VAST FS & Lustre FS \\
PFS R/W (GB/s) & 6.2 / 7.1 & 3.6 / 2.8 \\
\hline
\end{tabular}
\end{table}

\begin{table}[t]
\centering
\footnotesize
\setlength{\tabcolsep}{3pt}
\caption{Models used for evaluation. KV sizes are BF16 totals over all tensor-parallel ranks.}
\label{tab:models}
\begin{tabular}{p{27mm}|p{13mm}|p{13mm}|c}
\hline
\textbf{Model} & \textbf{KV/token} & \textbf{KV/chunk} & \textbf{Arch.} \\
 & \textbf{(KB)} & \textbf{(MB)} & \textbf{Ref.} \\
\hline
14B (Qwen3-14B) & 160 & 40 & \cite{Qwen/Qwen3-14B,qwen3_yang2025qwen3} \\
32B (DS-R1-Qwen-32B) & 256 & 64 & \cite{deepseek-ai/DeepSeek-R1-Distill-Qwen-32B,deepseek_r1_guo2025deepseek} \\
70B (DS-R1-Llama-70B) & 320 & 80 & \cite{deepseek-ai/DeepSeek-R1-Distill-Llama-70B,deepseek_r1_guo2025deepseek} \\
110B (Qwen1.5-110B) & 320 & 80 & \cite{Qwen/Qwen1.5-110B-Chat,qwen15_bai2023qwen} \\
\hline
\end{tabular}
\end{table}

\noindent\paragraph{\textbf{Models}} We evaluate four models from 14B to 110B parameters (Table~\ref{tab:models}). Their KV size grows from 160~KB per token at 14B to 320~KB at 70B and 110B, while compute grows with depth, so the four sizes span the balance between recomputing a prefix and reading it back. All models use GQA with 8 KV heads and a head dimension of 128 (compressing KV cache by up to 8$\times$ compared to standard MHA). We use TP=4 (2 KV heads per layer per GPU), and the default LMCache block size of 256 tokens produces 10--20~MB per GPU.

\noindent\paragraph{\textbf{Workloads}} We use three offline datasets and three online traces as detailed below, whose prefix reuse rate, is 80--96\%, and the study in \S~\ref{sec:eval:envelope} dilutes it down to 4\%. 

\begin{itemize}[topsep=0pt,itemsep=0pt,leftmargin=*]
\item \textit{LooGLE}~\cite{loogle_li2023loogle}: A long-document question-answering benchmark with documents spanning 2K--32K tokens. We use a merged subset consisting of long-dependency QA, short-dependency QA, and short-dependency cloze tasks, totaling $\sim$3.2K requests and $\sim$71M input tokens.

\item \textit{NarrativeQA}~\cite{narrativeqa_kocisky2018narrativeqa}: A reading comprehension benchmark, requiring understanding of entire narratives. We select a random subset of 2.5K queries, having an average of 46K~tokens/request. This is representative of very large context-lengths, producing 15K unique prefix blocks.

\item \textit{CUAD}~\cite{cuad_hendrycks2021cuad}: A dataset of legal contracts annotated for clause extraction. Each contract serves as a long-context prefix shared across multiple legal queries. We use 10K requests, averaging 11K tokens/request, producing 12K unique KV prefix blocks.
\end{itemize}

\noindent For the online tests, each request is sent at its scheduled arrival time, based on the following traces.

\begin{itemize}[topsep=0pt,itemsep=0pt,leftmargin=*]
\item \textit{Qwen-Coder}~\cite{kvcache_wang2025kvcache}: $\sim$7K requests from 716 code-generation sessions of a Qwen serving cluster on Alibaba Bailian.
Its output size averages to 109 tokens, and prompts average only 7.1K tokens, a hard case for storage offloading.

\item \textit{Chutes}~\cite{nixon2026year}: agentic sessions from a year-long production trace of the Chutes serving platform, with prompts of up to 130K tokens (62K on average), and outputs of 60 tokens.

\item \textit{SWE-bench agentic}~\cite{swebench_jimenez2024swebench}: 4000 requests from 130 coding-agent sessions resolving SWE-bench issues, with Poisson arrivals at 1.25~req/s. The average input and output token lengths for this trace averages at 19K and 95 tokens, respectively.
\end{itemize}

\noindent\paragraph{\textbf{Compared approaches}}
For all approaches, $\approx$320GB of host memory was used (80~GB/rank, and 40~GB/rank for Chutes traces outside the memory sweep of \S~\ref{sec:eval:envelope}), with the rest for compiled graphs, metadata management, coordination, and I/O buffers.
\begin{itemize}[topsep=0pt,itemsep=0pt,leftmargin=*,beginpenalty=10000]
\item \textit{CPU-only}: vLLM's prefix caching~\cite{APC} in spare GPU memory, extended by LMCache to host memory only~\cite{LMCache_CPURAM}. It is the baseline without storage offloading (recomputes on cache miss).
\item \textit{LMCache}: LMCache with host memory and node-local NVMe~\cite{cheng2025lmcache}. It flushes each KV chunk of each rank as one file, with multi-threaded, asynchronous \texttt{O\_DIRECT} I/O, and reads each chunk from the fastest tier that holds it. \proj extends this state-of-the-art connector.
\item \textit{Mooncake Store}: Mooncake's KV store~\cite{mooncake_qin2024mooncake}, used through Mooncake's own vLLM connector. Its memory pool and transfer buffers hold 80~GB per rank, and allows disk offloading only to a single tier, which in our case is the node-local NVMe, which is faster than PFS (Table~\ref{tab:testbeds}).
\item \textit{SGLang HiCache}: SGLang~\cite{sglang_zheng2024sglang} with its native radix-tree-based hierarchical cache, which keeps KV in 80~GB of host memory per rank and in files on the node-local NVMe (allows only a single storage tier). It represents a second serving engine with its own multi-tier cache. 
\item \proj: our engine, an LMCache backend (\S~\ref{sec:impl}) that reads from the local NVMe and the PFS.
\end{itemize}

\noindent\paragraph{\textbf{Key Evaluation Metrics}} For online traces, the primary metrics are TTFT (mean, p50, and p90) and SLO attainment, the fraction of \emph{offered} requests whose TTFT meets a deadline. Since our scope primarily applies for long-input workloads with few decodes (60--100 tokens in real-world traces outlined above), TTFT is the most prominent metrics from a user's perspective. For offline traces, the primary metric is throughput: the total tokens processed divided by the wall-clock time to serve the whole dataset. To explain the results, we also report per-step retrieve, compute, and store times, the read bandwidth of each tier, the storage read and write volume per rank, and the gap to the effective aggregate-bandwidth bound (\S~\ref{sec:model:bound}).

\begin{figure*}[!ht]
    \minipage[t]{0.32\textwidth}
        \centering
        \includegraphics[width=\linewidth]{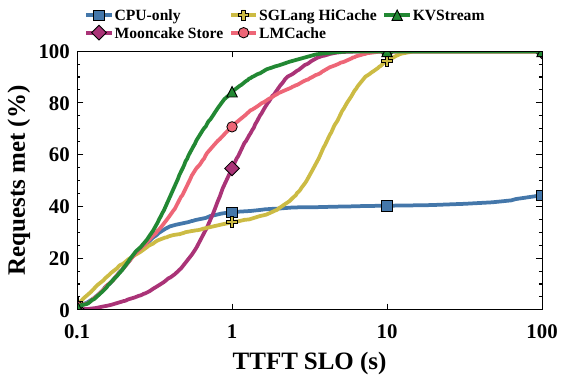}
        \captionsetup{skip=2pt,justification=raggedright}
        \caption{TTFT SLO attainment, first 3550 requests (SWE-bench agentic, 32B, \polaris).}
        \label{fig:o1-slo}
    \endminipage
    \hfill
    \minipage[t]{0.32\textwidth}
        \centering
        \includegraphics[width=\linewidth]{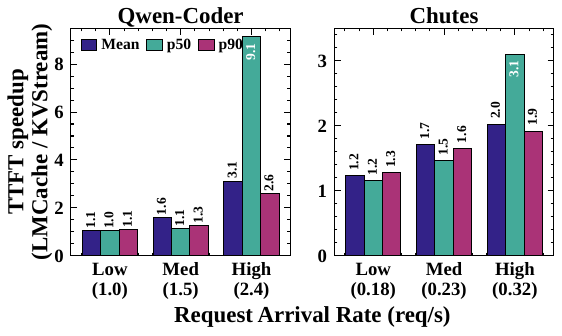}
        \captionsetup{skip=2pt,justification=raggedright}
        \caption{Mean, p50, and p90 TTFT speedup over LMCache vs.\ offered arrival rate (32B).}
        \label{fig:o6-load}
    \endminipage
    \hfill
    \minipage[t]{0.32\textwidth}
        \centering
        \includegraphics[width=\linewidth]{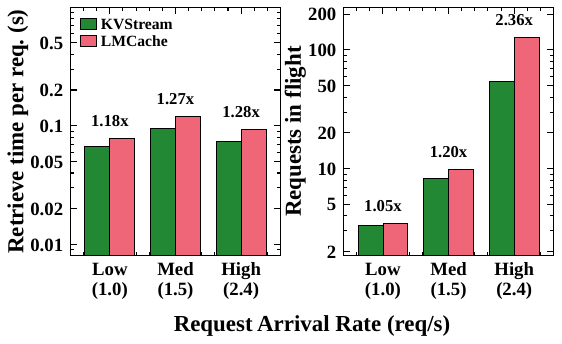}
        \captionsetup{skip=2pt,justification=raggedright}
        \caption{Retrieve time per request and requests in flight vs.\ arrival rate (Qwen-Coder).}
        \label{fig:o10-qwen}
    \endminipage
\end{figure*}

\noindent\paragraph{\textbf{Arrival rate factor}}\label{sec:eval:rates} To evaluate the request processing throughput of different approaches, we speed up or slow down the real-arrival timestamps of the Qwen-Coder and Chutes traces. This speed up factor does not change the order of request scheduling or arrival, and is only representative of under-vs-over-provisioned inference datacenters. We make three classification of this arrival rate factor based on the server's peak processing capacity when using \proj, which is 2.9 req/s on Qwen-Coder, and 0.46~req/s on Chutes. The load $\rho$ is the arrival rate divided by the node's capacity. A load of 33\% thus means that requests arrive at one third of the rate at which the node can serve them. We call a rate low below 35\% load, medium near 50\%, and high above 65\%. For the Qwen-Coder dataset (smaller in input length), the arrival rate for low, moderate, and high loads amount to 1.0~req/s, 1.5~req/s, and 2.4~req/s, respectively. For the Chutes dataset (high, 62K input tokens on average), the arrival rate for low, moderate, and high amounts to 0.18~req/s, 0.23~req/s, and 0.32~req/s, respectively. Chutes' arrivals are bursty, so its busiest 10-minute window reaches 54--107\% load. Waiting grows as $\rho/(1-\rho)$ (\S~\ref{sec:model:gains}), which is 0.5--0.6 at the low rates, 1.0 at the medium rates, and 2.2--4.7 at the high rates. At a low rate few requests queue, so TTFT shows only the per-request saving; at a high rate the queue dominates TTFT and amplifies savings across step.

\subsection{\textbf{Online Serving: TTFT, SLOs, Load, and Reuse}}
\label{sec:eval:online}
\label{sec:eval:queueing}
\label{sec:eval:sensitivity}

All online runs use one \polaris node, and measure the fraction of requests against different TTFT SLO(s) in Figure~\ref{fig:o1-slo}. Against, TTFT SLO of 1s, 84\% of requests meet the deadline with \proj, followed by 70\% with LMCache, 55\% with Mooncake Store, 37.7\% with CPU-only, and 34\% with SGLang's HiCache disk backend approach. SGLang HiCache is the fastest system while the working set fits in host memory (median TTFT 0.19~s over the first 1200 requests), but its file tier slows it once requests spill to storage. For the remainder of evaluations, we fix the serving engine to vLLM and vary other parameters. Furthermore, since LMCache has lower mean TTFT (1.15~s) and p50 (0.53~s) compared to Mooncake, which has mean TTFT of 1.18~s and p50 of 0.92s, we consider LMCache as the stronger state-of-the-art baseline in the remainder of evaluations. Appendix~\ref{sec:appendix:online} repeats this comparison in SGLang, where \proj beats HiCache and Mooncake (Figure~\ref{fig:sglang}).

In the next set of experiments in Figure~\ref{fig:o6-load}, we raise the arrival rate from low to medium to high (\S~\ref{sec:eval:rates}). On Qwen-Coder, \proj's mean-TTFT speedup over LMCache grows from 1.07$\times$ to 1.61$\times$ to 3.12$\times$, whereas on Chutes, the speedup grows from 1.23$\times$ to 1.71$\times$ to 2.01$\times$. To associate these gains to faster retrievals vs queueing delays, for the Qwen-Coder dataset, we observe in Figure~\ref{fig:o10-qwen}, per request, \proj spends 15--22\% less time retrieving than LMCache at every rate (0.067--0.095~s against 0.078--0.120~s). Meanwhile, the requests that wait behind each retrieval grow from 3.3 to 54.3 with \proj, and from 3.4 to 128 with LMCache. This demonstrates the queue, not the per-request saving, carries the growth. At low request arrival rate, the average in-flight request difference between \proj and LMCache is low (Figure~\ref{fig:o10-qwen} (right)), which amplifies with higher arrival rate, and despite nominal retrieval speedup (Figure~\ref{fig:o10-qwen} (left)), the gains amplify over the serving lifetime to achieve up to 3.12$\times$ faster TTFT compared to LMCache.

\begin{figure*}[!t]
    \minipage[t]{0.32\textwidth}
        \centering
        \includegraphics[width=\linewidth]{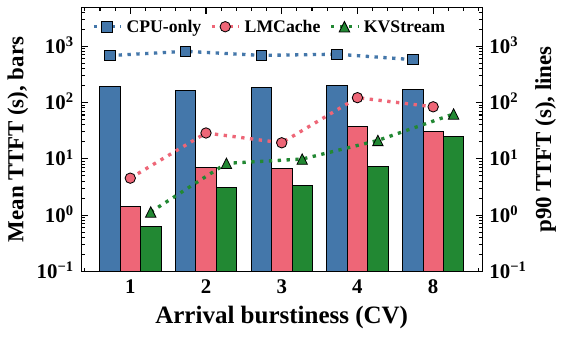}
        \captionsetup{skip=2pt,justification=raggedright}
        \caption{Mean (bars) and p90 (lines) TTFT vs.\ arrival burstiness (SWE-bench agentic).}
        \label{fig:o2-burst}
    \endminipage
    \hfill
    \minipage[t]{0.32\textwidth}
        \centering
        \includegraphics[width=\linewidth]{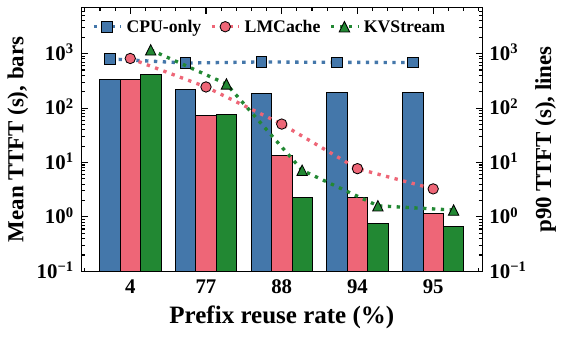}
        \captionsetup{skip=2pt,justification=raggedright}
        \caption{Mean (bars) and p90 (lines) TTFT vs.\ prefix reuse rate (SWE-bench agentic).}
        \label{fig:o3-reuse}
    \endminipage
    \hfill
    \minipage[t]{0.32\textwidth}
        \centering
        \includegraphics[width=\linewidth]{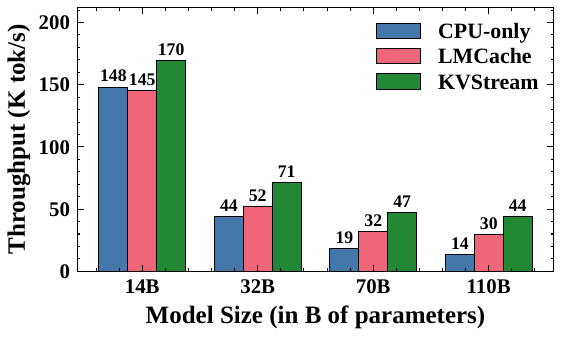}
        \captionsetup{skip=2pt,justification=raggedright}
        \caption{Throughput across model sizes (LooGLE, TP=4, \jlse).}
        \label{fig:model-throughput}
    \endminipage
\end{figure*}

In our next set of experiments, we study burst arrivals by varying the coefficient of variation~(CV) of the arrivals up to 8 (beyond the CV=4.3 of production arrivals~\cite{xiang2026servegen}) on the SWE-bench agentic traces. Despite variable CV, each run processes the same fixed set of 4000 requests, and their TTFT mean and p90 are reported in Figure~\ref{fig:o2-burst}. We observe that compared to LMCache, \proj's mean-TTFT advantage is 2.12$\times$ at an arrival CV of 1 and 6.07$\times$ at 4. It is 1.69--1.78$\times$ at 2--3, and 1.17$\times$ at 8 (the stress point beyond the production CV). We also observe that the TTFT p90 of \proj also consistently outperforms the compared approaches, absorbing burstiness impact. 

We then synthetically vary the prefix reuse volume of the SWE-bench traces by controlling the input prefix chain, and run all approaches for all 4000 requests. This studies (a) the impact of lower reuse volume $V$, described in Equation~\ref{eq:demand}, and (b) and higher write:read proportion, which in turn also includes the cost of write amplification due to multi-tier replication with \proj. Figure~\ref{fig:o3-reuse} dilutes the reuse of the SWE-bench trace with every other knob fixed. We observe that \proj's TTFT mean and p90 outperform both approaches, and against LMCache the advantage peaks at 6.0$\times$ at 88\% reuse. Below that, each rank writes more new KV while it reads less back: at 77\% reuse the two systems are on par (0.97$\times$) and at 4\% reuse \proj's speedup falls to 0.81$\times$, and the advantage falls to 0.8$\times$ compared to LMCache, showing \proj's failure envelope.

\begin{figure*}[!t]
    \minipage[t]{0.32\textwidth}
        \centering
        \includegraphics[width=\linewidth]{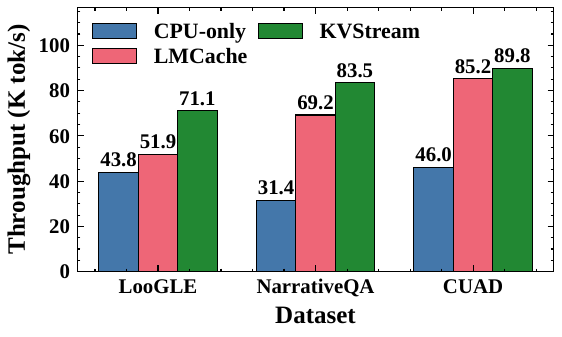}
        \captionsetup{skip=2pt,justification=raggedright}
        \caption{Throughput across datasets (32B, TP=4, \jlse).}
        \label{fig:dataset-diversity}
    \endminipage
    \hfill
    \minipage[t]{0.32\textwidth}
        \centering
        \includegraphics[width=\linewidth]{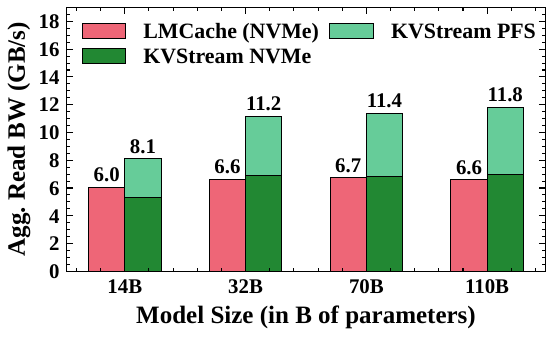}
        \captionsetup{skip=2pt,justification=raggedright}
        \caption{Read bandwidth by tier (LooGLE, TP=4, \jlse).}
        \label{fig:io-throughput}
    \endminipage
    \hfill
    \minipage[t]{0.32\textwidth}
        \centering
        \includegraphics[width=\linewidth]{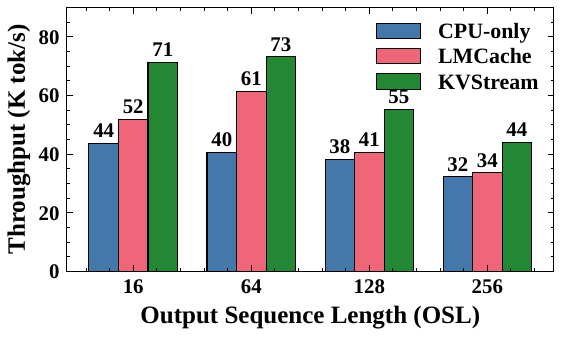}
        \captionsetup{skip=2pt,justification=raggedright}
        \caption{Throughput vs.\ output sequence length (32B, \jlse).}
        \label{fig:osl-scaling}
    \endminipage
\end{figure*}

\subsection{\textbf{Closed-Loop Offline Dataset Results}}
\label{sec:eval:offline}
\label{sec:eval:model-scaling}
\label{sec:eval:models}
\label{sec:eval:io-throughput}
\label{sec:eval:osl}

Closed-loop offline serving submit large batches of requests to the inference engine, which leads to queuing of all requests at once. Since offline serving keeps the engine saturated, we measure the peak processing throuhgputs (total tokens processed divided by total time) of different approaches (which does not include queueing delays). On one \jlse node (H100-80GB GPUs), \proj delivers higher throughput than both baselines at every model size (Figure~\ref{fig:model-throughput}). It reaches 1.17$\times$, 1.37$\times$, 1.48$\times$, and 1.49$\times$ the throughput of LMCache at 14B, 32B, 70B, and 110B, and 1.15$\times$ to 3.22$\times$ that of CPU-only. The advantage grows with model size for two reasons. A larger KV footprint per token leaves fewer prefixes in host memory (based on hardware, so same for all models), so more reads go to storage. And recomputation grows with model depth faster than KV grows: the 70B model is 5$\times$ larger than the 14B model but holds only 2$\times$ more KV per token (Table~\ref{tab:models}). The gain comes entirely from retrieval. LMCache and \proj spend nearly the same compute and store time at each size (Figure~\ref{fig:model-breakdown} in Appendix~\ref{sec:appendix:offline}), so replication adds no visible cost to an offline step.

Across datasets at 32B (Figure~\ref{fig:dataset-diversity}), \proj is 1.37$\times$, 1.21$\times$, and 1.05$\times$ faster than LMCache on LooGLE, NarrativeQA, and CUAD, and 1.62$\times$ to 2.66$\times$ faster than CPU-only. The gain is smallest on CUAD, whose requests are the shortest (11K tokens on average, against about 22K on LooGLE and 46K on NarrativeQA), consistent with a smaller retrieval share of its serving step (\S~\ref{sec:model:gains}). Figure~\ref{fig:io-throughput} shows why: \proj reads at the combined bandwidth of both tiers. Over the four ranks of a node, LMCache reads 6.0--6.7~GB/s from the NVMe. \proj reads about as much from the NVMe, adds 2.8--4.9~GB/s from the PFS, and reaches 8.1--11.8~GB/s in total. Both systems retrieve similar volume of prefixes from the storage tier (Figure~\ref{fig:tier-dist} in Appendix~\ref{sec:appendix}), attributing this speedup to faster parallel reads of \proj.

In our next set of experiments, we evaluate the effect of increasing output length on the throughput in Figure~\ref{fig:osl-scaling}. Given the auto-regressive, iterative nature of decode phases, we observe each step/request spends larger fraction of it's time on decodes, thereby diluting retrieval's share on the iteration. As modelled in the service time of \S~\ref{sec:model:gains}, the throughput speedup obtained by faster retrievals is therefore capped by Amdahl's law. We see in Figure~\ref{fig:osl-scaling} that on LooGLE and \jlse, from 16 to 256 output tokens, \proj's advantage over LMCache stays between 1.19$\times$ and 1.37$\times$.

\subsection{\textbf{Attaining the Aggregate-Bandwidth Bound}}
\label{sec:eval:validation}
\label{sec:eval:work-stealing}

In the next series of experiments, we test the two predictions of \S~\ref{sec:model:bound}: if completion-driven routing attains the aggregate-bandwidth bound $\underline{T}_i$ (Eq.~\ref{eq:bound}), and if a static split pays the penalty of Equation~\ref{eq:penalty}, which grows with the mismatch between its share $g$ and the bandwidth share $\varphi(i)$ (Eq.~\ref{eq:phi}).

\begin{figure*}[!t]
    \minipage[t]{0.32\textwidth}
        \centering
        \includegraphics[width=\linewidth]{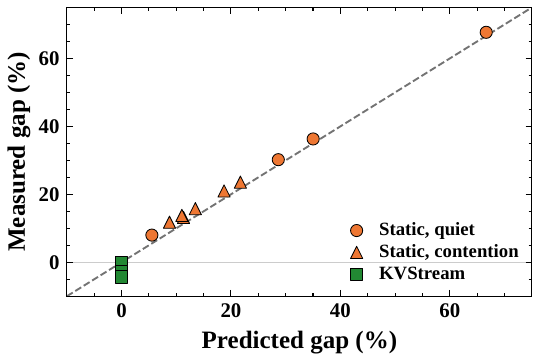}
        \captionsetup{skip=2pt,justification=raggedright}
        \caption{Measured vs.\ predicted gap to the bound, one point per run (32B, LooGLE, \polaris).}
        \label{fig:bound-accuracy}
    \endminipage
    \hfill
    \minipage[t]{0.32\textwidth}
        \centering
        \includegraphics[width=\linewidth]{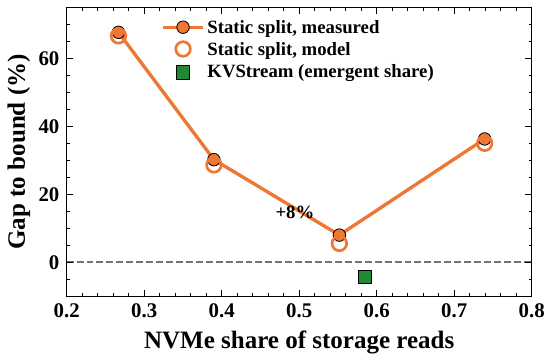}
        \captionsetup{skip=2pt,justification=raggedright}
        \caption{Gap to the bound vs.\ the NVMe share of storage reads on a quiet file system.}
        \label{fig:bound-share}
    \endminipage
    \hfill
    \minipage[t]{0.32\textwidth}
        \centering
        \includegraphics[width=\linewidth]{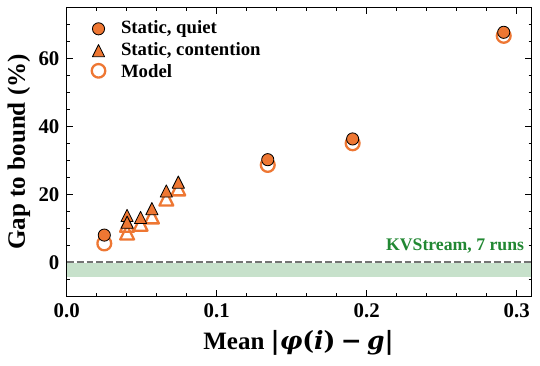}
        \captionsetup{skip=2pt,justification=raggedright}
        \caption{Gap to the bound vs.\ mean mismatch between $g$ and the bandwidth share $\varphi(i)$.}
        \label{fig:bound-mismatch}
    \endminipage
\end{figure*}

We use 17 runs of the 32B model on LooGLE on \polaris, and for every step, \emph{kvsim} computes the bound and $\varphi(i)$~(Eq.~\ref{eq:phi}) from the run's per-tier byte and time counters. On a quiet file system, four static placements put 60--90\% of each prefix's chunks on the NVMe. A static placement controls where chunks go, not where reads go: host memory serves the initial chunks of many prefix chain which reside on the NVMe, so the NVMe serves a smaller share $g$ of the storage demand: 27--74\% instead of 60--90\% which are stored on it. We first check the model run by run in Figure~\ref{fig:bound-accuracy}: Equation~\ref{eq:static} predicts the gap of every static run within 3\%, and \proj's runs sit within 4.3\% of the bound. The model thus explains the retrieval time of both policies, so the gap to the bound is a faithful measure of what each policy loses.

Next, Figure~\ref{fig:bound-share} sweeps the static share $g$ on the quiet file system. As Equation~\ref{eq:penalty} predicts, the gap is V-shaped in $g$: it is smallest (8.0\%) when $g$ matches the mean bandwidth share ($\approx 0.55$), and it grows on either side, up to 67.7\%. Even the best static split pays, because $\varphi(i)$ moves from step to step on a quiet file system too (O4), and a fixed $g$ matches it only on average. \proj has no share to tune, since completion-driven routing sets its split as reads complete, and it stays at the bound. End to end, \proj's throughput is 4.6\% above that of the best static split and 24.8\% above that of the worst, so even a well-tuned static split leaves throughput on the table.

For next experiments, a co-located \texttt{fio} job adds periodic, bursty, or steady traffic on the PFS, representative of other jobs sharing it. Figure~\ref{fig:bound-mismatch} shows that this traffic widens the spread of $\varphi(i)$~(Eq.~\ref{eq:phi}) and, for steady traffic, shifts its mean from 0.55 to 0.62, away from the static split's fixed $g = 0.55$. The static gap thus grows with the mean mismatch $|\varphi(i) - g|$, from 8.0\% on the quiet file system to 11.8--23.6\%, whereas \proj stays within 4.3\% of the bound under every pattern, because completion-driven routing rebalances the tiers at every step without monitoring their bandwidths (\S~\ref{sec:design:work-stealing}, M4).

\subsection{\textbf{Demand and Write Cost: Host Memory and Arrival Order}}
\label{sec:eval:envelope}
\label{sec:eval:workingset}
\label{sec:eval:order}
\label{sec:eval:replication}

\begin{figure*}[!t]
    \minipage[t]{0.32\textwidth}
        \centering
        \includegraphics[width=\linewidth]{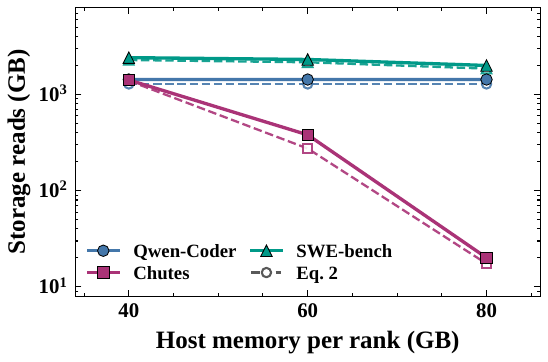}
        \captionsetup{skip=2pt,justification=raggedright}
        \caption{Storage reads per rank vs.\ host memory: measured and predicted by Equation~\ref{eq:demand}.}
        \label{fig:o12-reuse}
    \endminipage
    \hfill
    \minipage[t]{0.32\textwidth}
        \centering
        \includegraphics[width=\linewidth]{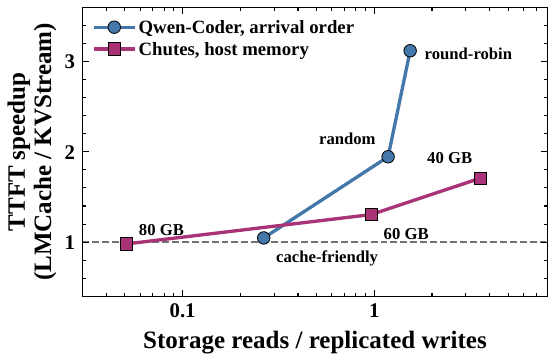}
        \captionsetup{skip=2pt,justification=raggedright}
        \caption{TTFT speedup over LMCache vs.\ storage reads relative to replicated writes (32B).}
        \label{fig:reads-writes}
    \endminipage
    \hfill
    \minipage[t]{0.32\textwidth}
        \centering
        \includegraphics[width=\linewidth]{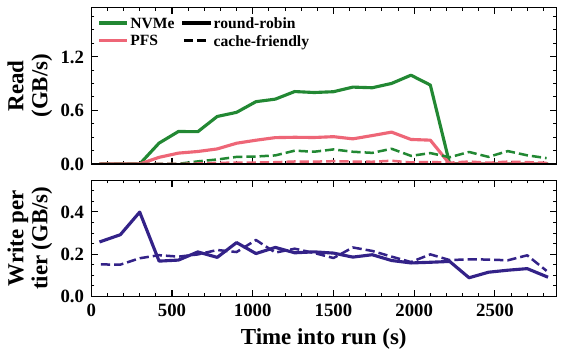}
        \captionsetup{skip=2pt,justification=raggedright}
        \caption{Per-rank read and write throughput of Qwen-Coder (high arrival rate, two orders).}
        \label{fig:io-online}
    \endminipage
\end{figure*}

In the next series of experiments, we test Equation~\ref{eq:demand}, which predicts the storage demand from the reuse distance $d$ and the cache capacity $M$, by varying host memory (40, 60, and 80~GB per rank) on the three online traces. Figure~\ref{fig:o12-reuse} shows that the prediction tracks the measured storage reads across all of them, from 20 to 2410~GB per rank (correlation 0.998). One rule produces three shapes, depending on where the reuse distances lie relative to $M$. On Qwen-Coder, they lie far beyond every memory size (median 278~GB per rank), so reads stay flat at 1428~GB per rank, and only faster retrieval helps. On SWE-bench, most session turns recur beyond the largest memory size, so reads decline only gradually (17\% from 40 to 80~GB). On Chutes, the reuse distances straddle the memory sizes, so storage reads drastically reduce-- 70$\times$ from 40 to 80~GB. The same rule holds across arrival orders: at the same memory, Qwen-Coder reads 5.8$\times$ more in round-robin order than in its cache-friendly order, and Equation~\ref{eq:demand} predicts each order within 26\%. More host memory therefore takes disk I/O off the critical path only when a workload's reuse distances straddle it, while long, interleaved sessions keep spilling. Appendix~\ref{sec:appendix:demand}, Figure~\ref{fig:o8-dram} extends this evaluation.

Replication, in contrast, writes every new chunk to both tiers, so its write volume is fixed by the workload (e.g., 1100~GB per rank for Qwen-Coder in every order), while its read benefit scales with the demand above. Figure~\ref{fig:reads-writes} therefore plots the TTFT speedup against the storage reads relative to the replicated writes. In both workloads, the speedup follows this ratio, as \S~\ref{sec:model:gains} predicts. On Qwen-Coder, moving from the cache-friendly to the random and round-robin orders raises reads from 0.27 to 1.18 and 1.54 of the writes, and the speedup from 1.05$\times$ to 1.95$\times$ and 3.12$\times$ (per order in Appendix~\ref{sec:appendix}, Figure~\ref{fig:o4-order}). On Chutes, more host memory lowers reads from 3.6 to 0.05 of the writes, and the speedup falls from 1.71$\times$ to 0.98$\times$. Multi-tier replication thus pays when storage reads dominate the replicated writes, the regime of interleaved multi-session serving that respects arrival order, and it stays within 2\% of single-tier offloading when the working set fits in host memory.

Deferred writes keep most of this write traffic off the read path (\S~\ref{sec:design:deferred-writes}, M2). On LooGLE, SWE-bench, and Chutes at 40--60~GB, writes overlap only 10--14\% of the NVMe's read time, because each forward pass is long enough to drain the step's writes (\S~\ref{sec:model:gains}). Qwen-Coder's short prompts leave less time, so its writes overlap 28--50\% of the NVMe's read time. Figure~\ref{fig:io-online} shows why order decides whether this cost pays: each tier writes 0.19~GB/s per rank in both orders, whereas reads peak at 1.27~GB/s in round-robin order and at only 0.21~GB/s in the cache-friendly order, where the writes nearly match the reads they serve. Figure~\ref{fig:o11-timeline} follows the round-robin run over time: the writes are exactly replicated, and the PFS serves close to its share of the read bandwidth (Appendix~\ref{sec:appendix:demand}).

\begin{figure*}[!t]
    \minipage[t]{0.32\textwidth}
        \centering
        \includegraphics[width=\linewidth]{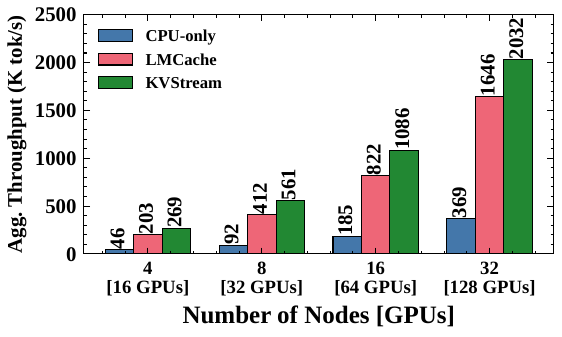}
        \captionsetup{skip=2pt,justification=raggedright}
        \caption{Throughput under data-parallel scaling to 32 nodes (NarrativeQA, 32B, \polaris).}
        \label{fig:dp-scaling}
    \endminipage
    \hfill
    \minipage[t]{0.32\textwidth}
        \centering
        \includegraphics[width=\linewidth]{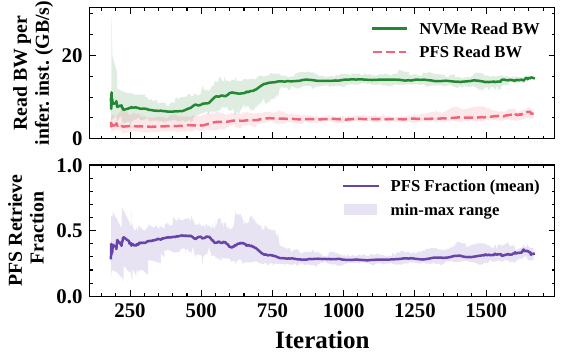}
        \captionsetup{skip=2pt,justification=raggedright}
        \caption{Per-replica read bandwidth and PFS share at 32 nodes: mean and min--max range.}
        \label{fig:cross-rank-variance}
    \endminipage
    \hfill
    \minipage[t]{0.32\textwidth}
        \centering
        \includegraphics[width=\linewidth]{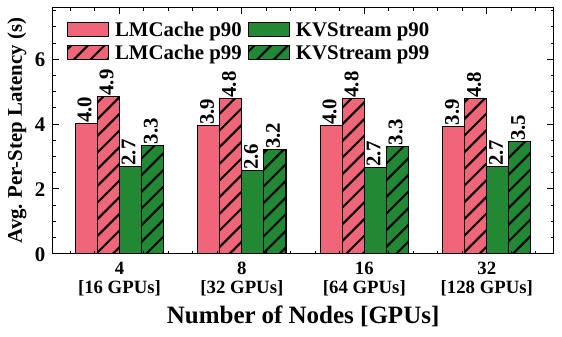}
        \captionsetup{skip=2pt,justification=raggedright}
        \caption{p90 and p99 step time under data-parallel scaling, averaged over replicas (\polaris).}
        \label{fig:latency-percentiles}
    \endminipage
\end{figure*}

\subsection{\textbf{Scaling and Ablations}}
\label{sec:eval:scaling}
\label{sec:eval:ablation}

In the next series of evaluations, we scale \proj out with data parallelism~(DP) from 4 to 32 \polaris nodes (16 to 128 GPUs) on NarrativeQA, where every replica has its own NVMe but all replicas share the PFS, the highest PFS concurrency in our evaluation. \proj keeps a 1.23--1.36$\times$ throughput advantage over LMCache (which operates on local NVMes, running in embarassingly parallel fashion, unaffected by other replicas) at every scale (Figure~\ref{fig:dp-scaling}), because each replica adapts to the contention on its own: at any iteration, some replicas read more than 50\% of their chunks from the PFS while others read less than 20\% (Figure~\ref{fig:cross-rank-variance}), without any coordination across nodes. Tail latency stays stable as well (Figure~\ref{fig:latency-percentiles}): \proj's p90 step time stays near 2.7~s at every scale, against 4.0~s for LMCache at 4 nodes, and even its p99 at 32 nodes remains 1.4$\times$ below LMCache's.

Table~\ref{tab:ablation} then adds the mechanisms of \S~\ref{sec:design} to LMCache one at a time. Offline, batched retrieval (M1) and deferred writes (M2) change throughput by less than 1\%, because a single NVMe caps the read bandwidth however its reads and writes are scheduled (O1, O3). The gain comes from the second tier: replicating every chunk on the PFS and reading a fixed 70:30 split (M3) raises throughput to 1.24$\times$ LMCache's, and completion-driven routing (M4) raises it to 1.48$\times$, because it sets the split as reads complete instead of fixing it (\S~\ref{sec:eval:validation}).

For online serving, at the high arrival rate, the queue amplifies these step-level savings (\S~\ref{sec:model:gains}): batched retrieval (M1) shortens the step by only 4.8\%, yet due to faster queue drains, it lowers mean TTFT 1.54$\times$, and deferred writes (M2) add little on top (1.58$\times$). A static split, in contrast, raises mean TTFT to 3.9$\times$ LMCache's. Each new turn of a session appends its chunks at the end of the prefix, so a split by prefix position lands most new KV on the PFS, which then serves 47\% of the reads with only 15\% of the read bandwidth; its reads take 3.2$\times$ the bound, close to the 3.1$\times$ that Equation~\ref{eq:penalty} predicts. As in \S~\ref{sec:eval:validation}, the placement controls where chunks go, not where reads go. With replication and completion-driven routing (M3 and M4), \proj's PFS share of reads (22\%) follows the PFS's share of the read bandwidth in that run (24\%), its reads take 1.02$\times$ the bound, and its mean TTFT is 2.39$\times$ below LMCache's. This further show that near saturation, accumulation of small changes in step time lead to large TTFT gains (\S~\ref{sec:model:gains}).

Finally, since full replication doubles the writes, we replay two replication policies in kvsim on the same traces based on the instantaneous per-tier per-step bandwidth and forward iteration times. With the first replication policy, we replicating chunks to the alternate storage tier after a tier turns slow, which leads to 1.19$\times$ write amplification instead of 2$\times$ write amplification in \proj, but it stays 12--74\% above the aggregate-bandwidth bound. This is because the repair (replication) arrives after the step has already stalled or suffered from slow reads. With the second policy, replicating a fixed 40\% of each prefix's chunks, leads to 1.43$\times$ write amplification but is up to 5\% slower than the aggregate-bandwidth bound of \proj.

\begin{table}[t]
\centering
\footnotesize
\setlength{\tabcolsep}{3.5pt}
\caption{Incremental ablation from LMCache to \proj. Offline: throughput, 70B, LooGLE, \jlse, M3 reads a fixed 70:30 NVMe:PFS split; Online: mean TTFT, 32B, Qwen-Coder in round-robin order at the high arrival rate, \polaris, M3 places each chunk on one tier, 80:20 by prefix position, without replication.}
\label{tab:ablation}
\begin{tabularx}{\linewidth}{X|X|X|X|X|X|X}
\hline
 & CPU-only & LMCache & + engine, & + deferred & + 2nd tier, & + completion- \\
 &  & (NVMe) & batched I/O & writes & static split & driven routing \\
 &  &  & (M1) & (M2) & (M3) & (M4) \\
\hline
Offline thruput (tok/s) [Higher better] & 18513 & 31977 & 32115 & 32177 & 39631 & \textbf{47210} \\
\quad vs.\ CPU-only & 1.00$\times$ & 1.73$\times$ & 1.73$\times$ & 1.74$\times$ & 2.14$\times$ & \textbf{2.55$\times$} \\
\hline
Online mean TTFT (s) [Lower better] & 340.7 & 31.1 & 20.2 & 19.8 & 122.1 & \textbf{13.0} \\
\quad vs.\ CPU-only & 1.00$\times$ & 10.9$\times$ & 16.9$\times$ & 17.3$\times$ & 2.79$\times$ & \textbf{26.2$\times$} \\
\hline
\end{tabularx}
\end{table}

\section{Beyond Our Testbeds: Hardware Trends, Other Tiers, and Composition}
\label{sec:discussion}

We now discuss how \proj's results carry beyond our testbeds, to larger memories, other tiers, and many nodes, and how it composes with orthogonal techniques: KV compression, request reordering, and disaggregated serving.

\paragraph*{\textbf{Larger memories}}
Larger nodes raise the cache capacity $M$ of Equation~\ref{eq:demand}, so reuses whose distance falls below $M$ stop reaching storage. A Vera Rubin node holds 288~GB of HBM and 750~GB of host memory per GPU~\cite{register2026rubin}, enough for most reuses of our round-robin Qwen-Coder trace, where \proj would gain little, as on Chutes at 80~GB (\S~\ref{sec:eval:envelope}). However, context lengths and session counts grow as well: a 1M-token session of a 70B model holds 320~GB of KV (Table~\ref{tab:models}), so a node that keeps 100 such sessions resumable needs about eight times the memory of a Vera Rubin node. Such workloads still spill, and the bandwidth of every storage path then sets their retrieval time.

\paragraph*{\textbf{Other second tiers}}
Equation~\ref{eq:bound} depends only on bandwidth, so any tier on a path independent of the node-local NVMe can serve as the second tier: a managed file system or object store attached to a cloud instance~\cite{aws_fsx_lustre,aws_s3}, remote NVMe, or CXL-attached memory. The added tier need not be as fast as the first, since completion-driven routing lets a slower tier simply claim fewer chunks. Its write path, however, must keep up with the replicated writes to beat single-tier systems (\S~\ref{sec:model:gains}).

\paragraph*{\textbf{Many nodes}}
Each node routes its own reads by completion, so nodes that share the PFS adapt without coordination: a node that finds the PFS congested reads more from its NVMe, which keeps \proj's advantage at up to 128 GPUs (\S~\ref{sec:eval:scaling}). Coordination could still save capacity and writes, for example by keeping one shared PFS copy of a prefix that several nodes reuse (particularly useful when dynamically scaling nodes up/down), which we leave to future work.

\paragraph*{\textbf{KV compression}}
Compressing KV by a factor $r$ divides both the reused volume and every reuse distance by $r$, so in Equation~\ref{eq:demand} it acts like $r$ times less volume and $r$ times more memory: $D_r = (V/r)\cdot\Pr[d > rM]$. Compression thus removes the demand of workloads whose reuse distances sit near memory capacity, but not the spill of long agentic and coding sessions whose working sets exceed memory by more than $r$: 100 resumable 128K-token sessions of a 70B model still need about 500~GB of KV after 8$\times$ compression, more than the 320~GB of host memory our nodes devote to KV. Compression and \proj therefore compose: the former shrinks what spills from memory, and the latter reads the spill at the aggregate bandwidth of multiple storage tiers.

\paragraph*{\textbf{Request reordering}}
Reuse-aware schedulers~\cite{sglang_zheng2024sglang,srivatsa2025preble,hu2026bidaw,xie2026peek} lower the demand of Equation~\ref{eq:demand} by moving the serving order toward a cache-friendly one, and \proj lowers the time per byte of whatever demand remains. Where the order is free, as in offline batches, the two compose, although in our cache-friendly order (similar schedule to BiDAW~\cite{hu2026bidaw}-- which lack public implementation) which such reordering approaches use, some KV reaches storage and a second tier adds little (\S~\ref{sec:eval:envelope}). Where it is not free to shuffle requests, as in online serving under per-request latency objectives or in multi-turn sessions whose turns depend on earlier outputs, the arrivals set the demand, and only the read path can be made faster. 

\paragraph*{\textbf{Disaggregated prefill and decode}}
Disaggregated serving runs prefill and decode on separate instances and moves KV between them through pooled memory and the network~\cite{distserve_zhong2024distserve,mooncake_qin2024mooncake,wang2025prefill,hong2025semi,feng2025windserve}. It changes where KV is produced and consumed, not how a prefix comes back from storage, so \proj's routing extends directly: a prefill step can read from the storage tiers or memory of both the prefill and the decode node as well as from the PFS, adding the bandwidth of each independent path to Equation~\ref{eq:bound}.

\section{Conclusions}
Once prefix KV outgrows GPU and host memory, storage bandwidth governs long-context LLM serving, yet the parallel file system or cloud file store that nodes already mount sits idle. We cast prefix retrieval as aggregate-bandwidth scheduling over the the available storage tiers. \proj batches a step's reads, defers writes behind the forward pass, replicates every chunk on every tier, and routes each read by completion. Our performance model shows that this attains the aggregate-bandwidth bound for any per-tier bandwidths, whereas any static split loses as the bandwidths drift or fluctuate, and that the end-to-end gain grows with the storage demand left by memory, arrival order, and with the load queued behind each read. In our evaluation on up to 128 GPUs, \proj retrieves within 5\% of the aggregate bandwidth bound, delivers up to 1.49$\times$ higher throughput than LMCache for offline/batched workloads, and lowers mean TTFT by up to 6.1$\times$ under bursty arrivals. Future work will reduce the replicated writes and add object stores as bandwidth sources.

\bibliographystyle{ACM-Reference-Format}
\bibliography{references}

@Misc{LustreFS,
  author = {{OpenSFS}},
  year = {2026},
  title = {Lustre: A parallel file system that supports many requirements of leadership class HPC simulation environments},
  howpublished = {\url{https://www.lustre.org/}},
  note = {Accessed: April 7, 2026}
}

@inproceedings{cacheblend,
  author = {Yao, Jiayi and Li, Hanchen and Liu, Yuhan and Ray, Siddhant and Cheng, Yihua and Zhang, Qizheng and Du, Kuntai and Lu, Shan and Jiang, Junchen},
  title = {{C}ache{B}lend: {F}ast {L}arge {L}anguage {M}odel {S}erving for {RAG} with {C}ached {K}nowledge {F}usion},
  year = {2025},
  url = {https://doi.org/10.1145/3689031.3696098},
  doi = {10.1145/3689031.3696098},
  booktitle = {Proceedings of the Twentieth European Conference on Computer Systems},
  pages = {94–109},
}

@article{cheng2025lmcache,
  title={{LMC}ache: {A}n {E}fficient {KV} {C}ache {L}ayer for {E}nterprise-{S}cale {LLM} {I}nference},
  author={Cheng, Yihua and Liu, Yuhan and Yao, Jiayi and An, Yuwei and Chen, Xiaokun and Feng, Shaoting and Huang, Yuyang and Shen, Samuel and Du, Kuntai and Jiang, Junchen},
  journal={arXiv preprint arXiv:2510.09665},
  year={2025},
  url={https://doi.org/10.48550/arXiv.2510.09665},
  doi={10.48550/arXiv.2510.09665},
}

@inproceedings{hu2026bidaw,
  title={{B}idaw: {E}nhancing $\{$Key-Value$\}$ {C}aching for {I}nteractive $\{$LLM$\}$ {S}erving via {B}idirectional $\{$Computation--Storage$\}$ {A}wareness},
  author={Hu, Shipeng and Zhang, Guangyan and Zhou, Yuqi and Wei, Yaya and Zhong, Ziyan and Chen, Jike},
  booktitle={24th USENIX Conference on File and Storage Technologies (FAST 26)},
  pages={101--116},
  year={2026},
  url={https://www.usenix.org/conference/fast26/presentation/hu-shipeng},
}

@online{APC,
  author = {vLLM},
  title = {{Automatic Prefix Caching - vLLM}},
  year = {2026},
  url = {https://docs.vllm.ai/en/stable/design/prefix_caching/},
  urldate = {2026-01-19},
  note = {Accessed: April 7, 2026}
}

@online{AWS_P5,
  author = {{Amazon Web Services}},
  title = {{Amazon EC2 P5 Instances – AWS}},
  year = {2026},
  url = {https://aws.amazon.com/ec2/instance-types/p5/},
  note = {Accessed: April 7, 2026}
}

@online{AWS_P4,
  author = {{Amazon Web Services}},
  title = {{Amazon EC2 P4d Instances – AWS}},
  year = {2026},
  url = {https://aws.amazon.com/ec2/instance-types/p4/},
  note = {Accessed: April 7, 2026}
}

@online{AWS_P6,
  author = {{Amazon Web Services}},
  title = {{Amazon EC2 P6e and P6 Instances - AWS}},
  year = {2026},
  url = {https://aws.amazon.com/ec2/instance-types/p6},
  note = {Accessed: October 8, 2026}
}

@online{LMCache_CPURAM,
  author = {LMCache},
  title = {{CPU RAM | LMCache}},
  year = {2026},
  url = {https://docs.lmcache.ai/kv_cache/storage_backends/cpu_ram.html},
  note = {Accessed: April 7, 2026}
}

@inproceedings{vllm_kwon2023efficient,
  title={{Efficient Memory Management for Large Language Model Serving with PagedAttention}},
  author={Woosuk Kwon and Zhuohan Li and Siyuan Zhuang and Ying Sheng and Lianmin Zheng and Cody Hao Yu and Joseph E. Gonzalez and Hao Zhang and Ion Stoica},
  booktitle={Proceedings of the ACM SIGOPS 29th Symposium on Operating Systems Principles},
  year={2023},
  doi={10.1145/3600006.3613165},
  url={https://doi.org/10.1145/3600006.3613165}
}

@article{mooncake_qin2024mooncake,
      title={{M}ooncake: {A} {KVC}ache-centric {D}isaggregated {A}rchitecture for {LLM} {S}erving},
  author={Qin, Ruoyu and Li, Zheming and He, Weiran and Cui, Jialei and Tang, Heyi and Ren, Feng and Ma, Teng and Cai, Shangming and Zhang, Yineng and Zhang, Mingxing and others},
  journal={ACM Transactions on Storage},
  year={2024},
  publisher={ACM New York, NY},
  doi={10.1145/3773772},
  url={https://doi.org/10.1145/3773772}
}

@inproceedings{kvcache_wang2025kvcache,
  title={$\{$KVCache$\}$ Cache in the Wild: Characterizing and Optimizing $\{$KVCache$\}$ Cache at a Large Cloud Provider},
  author={Wang, Jiahao and Han, Jinbo and Wei, Xingda and Shen, Sijie and Zhang, Dingyan and Fang, Chenguang and Chen, Rong and Yu, Wenyuan and Chen, Haibo},
  booktitle={2025 USENIX Annual Technical Conference (USENIX ATC 25)},
  pages={465--482},
  year={2025}
}

@article{computeorstorage_jin2024compute,
  title={Compute or load kv cache? why not both?},
  author={Jin, Shuowei and Liu, Xueshen and Zhang, Qingzhao and Mao, Z Morley},
  journal={arXiv preprint arXiv:2410.03065},
  year={2024},
  doi = {10.48550/arXiv.2410.03065},
  url = {https://doi.org/10.48550/arXiv.2410.03065},
}

@inproceedings{flexgen_sheng2023flexgen,
  title={{FlexGen: High-Throughput Generative Inference of Large Language Models with a Single GPU}},
  author={Sheng, Ying and Zheng, Lianmin and Yuan, Binhang and Li, Zhuohan and Ryabinin, Max and Chen, Beidi and Liang, Percy and R{\'e}, Christopher and Stoica, Ion and Zhang, Ce},
  booktitle={International Conference on Machine Learning},
  pages={31094--31116},
  year={2023},
  organization={PMLR},
  url={https://proceedings.mlr.press/v202/sheng23a.html}
}

@article{sglang_zheng2024sglang,
  title={{SGLang: efficient execution of structured language model programs}},
  author={Zheng, Lianmin and Yin, Liangsheng and Xie, Zhiqiang and Sun, Chuyue and Huang, Jeff and Yu, Cody H and Cao, Shiyi and Kozyrakis, Christos and Stoica, Ion and Gonzalez, Joseph E and others},
  journal={Advances in neural information processing systems},
  volume={37},
  pages={62557--62583},
  year={2024},
  doi={},
  url={https://dl.acm.org/doi/10.5555/3737916.3739916}
}

@inproceedings{cachegen_liu2024cachegen,
  title={{C}ache{G}en: {KV} {C}ache {C}ompression and {S}treaming for {F}ast {L}arge {L}anguage {M}odel {S}erving},
  author={Liu, Yuhan and Li, Hanchen and Cheng, Yihua and Ray, Siddhant and Huang, Yuyang and Zhang, Qizheng and Du, Kuntai and Yao, Jiayi and Lu, Shan and Ananthanarayanan, Ganesh and others},
  booktitle={Proceedings of the ACM SIGCOMM 2024 Conference},
  pages={38--56},
  year={2024},
  doi = {10.1145/3651890.3672274},
  url = {https://doi.org/10.1145/3651890.3672274},
}

@article{shadowkv_sun2024shadowkv,
  title={Shadowkv: Kv cache in shadows for high-throughput long-context llm inference},
  author={Sun, Hanshi and Chang, Li-Wen and Bao, Wenlei and Zheng, Size and Zheng, Ningxin and Liu, Xin and Dong, Harry and Chi, Yuejie and Chen, Beidi},
  journal={arXiv preprint arXiv:2410.21465},
  year={2024},
  doi = {10.48550/arXiv.2410.21465},
  url = {https://doi.org/10.48550/arXiv.2410.21465},
}

@inproceedings{infinigen_lee2024infinigen,
  title={$\{$InfiniGen$\}$: Efficient generative inference of large language models with dynamic $\{$KV$\}$ cache management},
  author={Lee, Wonbeom and Lee, Jungi and Seo, Junghwan and Sim, Jaewoong},
  booktitle={18th USENIX Symposium on Operating Systems Design and Implementation (OSDI 24)},
  pages={155--172},
  year={2024},
  doi={10.5555/3691938.3691947},
  url={https://dl.acm.org/doi/10.5555/3691938.3691947}
}

@article{loogle_li2023loogle,
  title={{LooGLE: Can Long-Context Language Models Understand Long Contexts?}},
  author={Li, Jiaqi and Wang, Mengmeng and Zheng, Zilong and Zhang, Muhan},
  journal={arXiv preprint arXiv:2311.04939},
  year={2023},
  doi={10.48550/arXiv.2311.04939},
  url = {https://doi.org/10.48550/arXiv.2311.04939},
}

@article{narrativeqa_kocisky2018narrativeqa,
    title = "The {N}arrative{QA} {R}eading {C}omprehension {C}hallenge",
    author = "Ko{\v{c}}isk{\'y}, Tom{\'a}{\v{s}}  and
      Schwarz, Jonathan  and
      Blunsom, Phil  and
      Dyer, Chris  and
      Hermann, Karl Moritz  and
      Melis, G{\'a}bor  and
      Grefenstette, Edward",
    editor = "Lee, Lillian  and
      Johnson, Mark  and
      Toutanova, Kristina  and
      Roark, Brian",
    journal = "Transactions of the Association for Computational Linguistics",
    volume = "6",
    year = "2018",
    address = "Cambridge, MA",
    publisher = "MIT Press",
    url = "https://aclanthology.org/Q18-1023",
    doi = "10.1162/tacl_a_00023",
    pages = "317--328"
}

@inproceedings{cuad_hendrycks2021cuad,
  author = {Hendrycks, Dan and Burns, Collin and Chen, Anya and Ball, Spencer},
 booktitle = {Proceedings of the Neural Information Processing Systems Track on Datasets and Benchmarks},
 editor = {J. Vanschoren and S. Yeung},
 pages = {},
 title = {{C}UAD: {A}n {E}xpert-{A}nnotated {NLP} {D}ataset for {L}egal {C}ontract {R}eview},
 url = {https://datasets-benchmarks-proceedings.neurips.cc/paper_files/paper/2021/file/6ea9ab1baa0efb9e19094440c317e21b-Paper-round1.pdf},
 volume = {1},
 year = {2021}
}

@article{qwen3_yang2025qwen3,
  title={{Q}wen3 {T}echnical {R}eport},
  author={Yang, An and Yang, Baosong and Zhang, Beichen and others},
  journal={arXiv preprint arXiv:2505.09388},
  year={2025},
  url={https://doi.org/10.48550/arXiv.2505.09388},
  doi={10.48550/arXiv.2505.09388}
}

@online{Qwen/Qwen3-14B,
  title = {Qwen/Qwen1.5-110B-Chat · Hugging Face},
  author = {Qwen},
  url = {https://huggingface.co/Qwen/Qwen3-14B},
  year = {2025},
  note = {Accessed: April 8, 2026}
}

@article{deepseek_r1_guo2025deepseek,
  title={{DeepSeek-R1: Incentivizing Reasoning Capability in LLMs via Reinforcement Learning}},
  author={Guo, Daya and Yang, Dejian and Zhang, He and others},
  journal={arXiv preprint arXiv:2501.12948},
  year={2025},
  url={https://doi.org/10.48550/arXiv.2501.12948},
  doi={10.48550/arXiv.2501.12948}
}

@online{deepseek-ai/DeepSeek-R1-Distill-Qwen-32B,
  title = {deepseek-ai/DeepSeek-R1-Distill-Qwen-32B · Hugging Face},
  author = {{deepseek-ai}},
  url = {https://huggingface.co/deepseek-ai/DeepSeek-R1-Distill-Qwen-32B},
  year = {2025},
  note = {Accessed: April 8, 2026}
}

@online{deepseek-ai/DeepSeek-R1-Distill-Llama-70B,
  title = {deepseek-ai/DeepSeek-R1-Distill-Llama-70B · Hugging Face},
  author = {{deepseek-ai}},
  url = {https://huggingface.co/deepseek-ai/DeepSeek-R1-Distill-Llama-70B},
  year = {2025},
  note = {Accessed: April 8, 2026}
}

@article{qwen15_bai2023qwen,
  title={{Q}wen {T}echnical {R}eport},
  author={Bai, Jinze and Bai, Shuai and Chu, Yunfei and others},
  journal={arXiv preprint arXiv:2309.16609},
  year={2023},
  url={https://doi.org/10.48550/arXiv.2309.16609},
  doi={10.48550/arXiv.2309.16609}
}

@online{Qwen/Qwen1.5-110B-Chat,
  title = {Qwen/Qwen1.5-110B-Chat · Hugging Face},
  author = {Qwen},
  url = {https://huggingface.co/Qwen/Qwen1.5-110B-Chat},
  year = {2024},
  note = {Accessed: April 8, 2026}
}

@article{wang2025prefill,
  title={Prefill-decode aggregation or disaggregation? unifying both for goodput-optimized llm serving},
  author={Wang, Chao and Zuo, Pengfei and Chen, Zhangyu and Liang, Yunkai and Yu, Zhou and Yang, Ming-Chang},
  journal={arXiv preprint arXiv:2508.01989},
  year={2025},
  url={https://doi.org/10.48550/arXiv.2508.01989},
  doi={10.48550/arXiv.2508.01989}
}

@article{hong2025semi,
  title={{semi-PD: Towards Efficient LLM Serving via Phase-Wise Disaggregated Computation and Unified Storage}},
  author={Hong, Ke and Chen, Lufang and Wang, Zhong and Li, Xiuhong and Mao, Qiuli and Ma, Jianping and Xiong, Chao and Wu, Guanyu and Han, Buhe and Dai, Guohao and others},
  journal={arXiv preprint arXiv:2504.19867},
  year={2025},
  url={https://doi.org/10.48550/arXiv.2504.19867},
  doi={10.48550/arXiv.2504.19867}
}

@inproceedings{feng2025windserve,
  title={{WindServe: Efficient Phase-Disaggregated LLM Serving with Stream-based Dynamic Scheduling}},
  author={Feng, Jingqi and Huang, Yukai and Zhang, Rui and Liang, Sicheng and Yan, Ming and Wu, Jie},
  booktitle={Proceedings of the 52nd Annual International Symposium on Computer Architecture},
  pages={1283--1295},
  year={2025},
  url={https://doi.org/10.1145/3695053.3730999},
  doi={10.1145/3695053.3730999}
}

@inproceedings{an2024fireflyer,
  title     = {Fire-Flyer {AI-HPC}: A Cost-Effective Software-Hardware Co-Design for Deep Learning},
  author    = {Wei An and Xiao Bi and Guanting Chen and Shanhuang Chen and Chengqi Deng and Honghui Ding and Kai Dong and Qiushi Du and Wenjun Gao and Kang Guan and Jianzhong Guo and Yongqiang Guo and Zhe Fu and Ying He and Panpan Huang and Jiashi Li and Wenfeng Liang and Xiaodong Liu and Xin Liu and Yiyuan Liu and Yuxuan Liu and Shanghao Lu and Xuan Lu and Xiaotao Nie and Tian Pei and Junjie Qiu and Hui Qu and Zehui Ren and Zhangli Sha and Xuecheng Su and Xiaowen Sun and Yixuan Tan and Min Tang and Shiyu Wang and Yaohui Wang and Yongji Wang and Ziwei Xie and Yi Xiong and Yanhong Xu and Shengfeng Ye and Shuiping Yu and Yukun Zhang and Liyue Zhang and Haowei Zhang and Mingchuan Zhang and Wentao Zhang and Yichao Zhang and Chenggang Zhao and Yao Zhao and Shangyan Zhou and Shunfeng Zhou and Yuheng Zou},
  booktitle = {SC24: International Conference for High Performance Computing, Networking, Storage and Analysis},
  pages     = {1314--1327},
  year      = {2024},
  publisher = {IEEE},
  url       = {https://doi.org/10.1109/SC41406.2024.00089},
  doi       = {10.1109/SC41406.2024.00089}
}

@article{li2025hotprefix,
  title     = {{HotPrefix: Hotness-Aware KV Cache Scheduling for Efficient Prefix Sharing in LLM Inference Systems}},
  author    = {Yuhang Li and Rong Gu and Chengying Huan and Zhibin Wang and Renjie Yao and Chen Tian and Guihai Chen},
  journal   = {Proceedings of the ACM on Management of Data},
  volume    = {3},
  number    = {4},
  articleno = {250},
  pages     = {1--27},
  year      = {2025},
  publisher = {ACM},
  url       = {https://doi.org/10.1145/3749168},
  doi       = {10.1145/3749168}
}

@inproceedings{srivatsa2025preble,
  title     = {{Preble: Efficient Distributed Prompt Scheduling for LLM Serving}},
  author    = {Vikranth Srivatsa and Zijian He and Reyna Abhyankar and Dongming Li and Yiying Zhang},
  booktitle = {The Thirteenth International Conference on Learning Representations (ICLR)},
  year      = {2025},
  url       = {https://openreview.net/pdf?id=meKEKDhdnx}
}

@inproceedings{gao2025fast,
  title={{Fast State Restoration in LLM Serving with HCache}},
  author={Gao, Shiwei and Chen, Youmin and Shu, Jiwu},
  booktitle={Proceedings of the Twentieth European Conference on Computer Systems},
  pages={128--143},
  year={2025},
  url={https://doi.org/10.1145/3689031.3696072},
  doi={10.1145/3689031.3696072}
}

@inproceedings{ren2025characterizing,
  title={{An I/O Characterizing Study of Offloading LLM Models and KV Caches to NVMe SSD}},
  author={Ren, Zebin and Doekemeijer, Krijn and De Matteis, Tiziano and Pinto, Christian and Stoica, Radu and Trivedi, Animesh},
  booktitle={Proceedings of the 5th Workshop on Challenges and Opportunities of Efficient and Performant Storage Systems},
  pages={23--33},
  year={2025},
  url={https://doi.org/10.1145/3719330.3721230},
  doi={10.1145/3719330.3721230}
}

@inproceedings{gqa_ainslie2023gqa,
  title={{GQA}: {Training Generalized Multi-Query Transformer Models from Multi-Head Checkpoints}},
  author={Ainslie, Joshua and Lee-Thorp, James and de Jong, Michiel and Zemlyanskiy, Yury and Lebr{\'o}n, Federico and Sanghai, Sumit},
  booktitle={Proceedings of the 2023 Conference on Empirical Methods in Natural Language Processing (EMNLP)},
  pages={4895--4901},
  year={2023},
  url={https://aclanthology.org/2023.emnlp-main.298/},
  doi={10.18653/v1/2023.emnlp-main.298}
}

@inproceedings{he2017unwritten,
  title={{The Unwritten Contract of Solid State Drives}},
  author={He, Jun and Kannan, Sudarsun and Arpaci-Dusseau, Andrea C. and Arpaci-Dusseau, Remzi H.},
  booktitle={Proceedings of the Twelfth European Conference on Computer Systems (EuroSys)},
  pages={127--144},
  year={2017},
  doi={10.1145/3064176.3064187},
  url={https://doi.org/10.1145/3064176.3064187}
}

@inproceedings{yan2017tinytail,
  title={{Tiny-Tail Flash: Near-Perfect Elimination of Garbage Collection Tail Latencies in NAND SSDs}},
  author={Yan, Shiqin and Li, Huaicheng and Hao, Mingzhe and Tong, Michael Hao and Sundararaman, Swaminathan and Chien, Andrew A. and Gunawi, Haryadi S.},
  booktitle={15th USENIX Conference on File and Storage Technologies (FAST)},
  pages={15--28},
  year={2017},
  url={https://www.usenix.org/conference/fast17/technical-sessions/presentation/yan}
}

@inproceedings{yildiz2016rootcauses,
  title={{On the Root Causes of Cross-Application I/O Interference in {HPC} Storage Systems}},
  author={Yildiz, Orcun and Dorier, Matthieu and Ibrahim, Shadi and Ross, Robert and Antoniu, Gabriel},
  booktitle={IEEE International Parallel and Distributed Processing Symposium (IPDPS)},
  pages={750--759},
  year={2016},
  doi={10.1109/IPDPS.2016.50},
  url={https://doi.org/10.1109/IPDPS.2016.50}
}

@article{attention_vaswani2017attention,
  title={{A}ttention is {A}ll {Y}ou {N}eed},
  author={Vaswani, Ashish and Shazeer, Noam and Parmar, Niki and Uszkoreit, Jakob and Jones, Llion and Gomez, Aidan N and Kaiser, {\L}ukasz and Polosukhin, Illia},
  journal={Advances in neural information processing systems},
  volume={30},
  year={2017},
  doi={},
  url={https://proceedings.neurips.cc/paper/2017/hash/3f5ee243547dee91fbd053c1c4a845aa-Abstract.html}
}

@inproceedings{attentioncompcomplexity_keles2023computational,
  title={{O}n {T}he {C}omputational {C}omplexity of {S}elf-{A}ttention},
  author={Keles, Feyza Duman and Wijewardena, Pruthuvi Mahesakya and Hegde, Chinmay},
  booktitle={International conference on algorithmic learning theory},
  pages={597--619},
  year={2023},
  organization={PMLR},
  doi={},
  url={https://proceedings.mlr.press/v201/duman-keles23a.html}
}

@article{llama3_grattafiori2024llama,
  title={{T}he {L}lama 3 {H}erd of {M}odels},
  author={Grattafiori, Aaron and Dubey, Abhimanyu and Jauhri, Abhinav and Pandey, Abhinav and Kadian, Abhishek and Al-Dahle, Ahmad and Letman, Aiesha and Mathur, Akhil and Schelten, Alan and Vaughan, Alex and others},
  journal={arXiv preprint arXiv:2407.21783},
  year={2024},
  doi={10.48550/arXiv.2407.21783},
  url={https://doi.org/10.48550/arXiv.2407.21783}
}

@article{contextlengthsummary_hsieh2024ruler,
  title={{RULER}: {W}hat's the {R}eal {C}ontext {S}ize of {Y}our {L}ong-{C}ontext {L}anguage {M}odels?},
  author={Hsieh, Cheng-Ping and Sun, Simeng and Kriman, Samuel and Acharya, Shantanu and Rekesh, Dima and Jia, Fei and Zhang, Yang and Ginsburg, Boris},
  journal={arXiv preprint arXiv:2404.06654},
  year={2024},
  doi={10.48550/arXiv.2404.06654},
  url={https://doi.org/10.48550/arXiv.2404.06654}
}

@article{llmapplication_liu2025comprehensive,
  title={{A} {C}omprehensive {S}urvey on {L}ong {C}ontext {L}anguage {M}odeling},
  author={Liu, Jiaheng and Zhu, Dawei and Bai, Zhiqi and He, Yancheng and Liao, Huanxuan and Que, Haoran and Wang, Zekun and Zhang, Chenchen and Zhang, Ge and Zhang, Jiebin and others},
  journal={arXiv preprint arXiv:2503.17407},
  year={2025},
  url={https://doi.org/10.48550/arXiv.2503.17407},
  doi={10.48550/arXiv.2503.17407}
}

@article{VELOC-FGCS24,
  title = {{Scalable I/O aggregation for asynchronous multi-level checkpointing}},
  journal = {Future Generation Computer Systems},
  volume = {160},
  pages = {420-432},
  year = {2024},
  issn = {0167-739X},
  doi = {10.1016/j.future.2024.06.003},
  url = {https://www.sciencedirect.com/science/article/pii/S0167739X24002929},
  author = {Gossman, Mikaila J. and Nicolae, Bogdan and Calhoun, Jon C.}
}

@inproceedings{KVCacheProfile-IPDPS25,
  title = {{Characterizing the Behavior and Impact of KV Caching on Transformer Inferences under Concurrency}},
  author = {Ye, Jie and Cernuda, Jaime and Maurya, Avinash and Sun, Xian-He and Kougas, Anthony and Nicolae, Bogdan},
  booktitle = {IPDPS'25: The 39th IEEE International Parallel and Distributed Processing Symposium},
  year = {2025},
  address = {Milan, Italy},
  url = {https://hal.inria.fr/hal-04984000},
  doi = {10.1109/IPDPS64566.2025.00108}
}

@misc{nvidia_dynamo_kvbm_2026,
  author       = {{NVIDIA Corporation}},
  title        = {KVBM Overview (v0.8.1)},
  howpublished = {NVIDIA Dynamo Documentation},
  year         = {2026},
  url          = {https://docs.nvidia.com/dynamo/v-0-8-1/components/kvbm/overview}
}

@techreport{ibm_redbooks_md260021_2026,
  author       = {{IBM Redbooks}},
  title        = {Context Without Limits: A High-Performance KV Cache Platform for Large-Scale AI Inference},
  institution  = {IBM},
  year         = {2026},
  url          = {https://www.redbooks.ibm.com/docs/MD260021/MD260021.html}
}

@article{nixon2026year,
  title={A Year in LLM Serving: Workload Evolution, Caching and Load-Balancing},
  author={Nixon, William and Durbin, Jon and Standhartinger, Florian and Gunawi, Haryadi S and Yang, Juncheng},
  journal={arXiv preprint arXiv:2608.13573},
  year={2026}
}

@inproceedings{h2o_zhang2023h2o,
  title     = {{H2O}: Heavy-Hitter Oracle for Efficient Generative Inference of Large Language Models},
  author    = {Zhang, Zhenyu and Sheng, Ying and Zhou, Tianyi and Chen, Tianlong and Zheng, Lianmin and Cai, Ruisi and Song, Zhao and Tian, Yuandong and R{\'e}, Christopher and Barrett, Clark and Wang, Zhangyang and Chen, Beidi},
  booktitle = {Advances in Neural Information Processing Systems (NeurIPS)},
  year      = {2023}
}

@inproceedings{streamingllm_xiao2024efficient,
  title     = {Efficient Streaming Language Models with Attention Sinks},
  author    = {Xiao, Guangxuan and Tian, Yuandong and Chen, Beidi and Han, Song and Lewis, Mike},
  booktitle = {International Conference on Learning Representations (ICLR)},
  year      = {2024}
}

@article{deepseekv2_2024,
  title   = {{DeepSeek-V2}: A Strong, Economical, and Efficient Mixture-of-Experts Language Model},
  author  = {{DeepSeek-AI}},
  journal = {arXiv preprint arXiv:2405.04434},
  year    = {2024}
}

@article{pan2025kvflow,
  title   = {{KVFlow}: Efficient Prefix Caching for Accelerating {LLM}-Based Multi-Agent Workflows},
  author  = {Pan, Zaifeng and Patel, Ajjkumar and Hu, Zhengding and Shen, Yipeng and Guan, Yue and Li, Wan-Lu and Qin, Lianhui and Wang, Yida and Ding, Yufei},
  journal = {arXiv preprint arXiv:2507.07400},
  year    = {2025}
}

@article{ouyang2026unicache,
  title   = {{UniCache}: Unifying Prefix Cache Eviction for Heterogeneous {LLM} Serving Workloads},
  author  = {Ouyang, Bei and Qiao, Yifan and Xing, Jiarong},
  journal = {Proceedings of the ACM on Measurement and Analysis of Computing Systems},
  volume  = {10},
  number  = {2},
  articleno = {54},
  year    = {2026},
  doi     = {10.1145/3805652}
}

@article{xie2026peek,
  title   = {{PEEK}: Predictive Queue-Informed {KV} Cache Management for {LLM} Serving},
  author  = {Xie, Bing and Wang, Zhipeng and Tanaka, Masahiro and Zheng, Zhen},
  journal = {arXiv preprint arXiv:2607.02525},
  year    = {2026}
}

@article{kanichai2026pykvcache,
  title   = {Building py-kvcache: A Performance Characterization of External {KV} Caching for {vLLM} with {NVMe} {SSDs}},
  author  = {Kanichai, Joseph and De Matteis, Tiziano and Trivedi, Animesh},
  journal = {arXiv preprint arXiv:2609.11744},
  year    = {2026}
}

@inproceedings{burstgpt_wang2025burstgpt,
  title     = {{BurstGPT}: A Real-World Workload Dataset to Optimize {LLM} Serving Systems},
  author    = {Wang, Yuxin and Chen, Yuhan and Li, Zeyu and Kang, Xueze and Fang, Yuchu and Zhou, Yeju and Zheng, Yang and Tang, Zhenheng and He, Xin and Guo, Rui and Wang, Xin and Wang, Qiang and Zhou, Amelie Chi and Chu, Xiaowen},
  booktitle = {Proceedings of the 31st ACM SIGKDD Conference on Knowledge Discovery and Data Mining (KDD)},
  year      = {2025},
  doi       = {10.1145/3711896.3737413}
}

@inproceedings{swebench_jimenez2024swebench,
  title     = {{SWE-bench}: Can Language Models Resolve Real-World {GitHub} Issues?},
  author    = {Jimenez, Carlos E. and Yang, John and Wettig, Alexander and Yao, Shunyu and Pei, Kexin and Press, Ofir and Narasimhan, Karthik},
  booktitle = {International Conference on Learning Representations (ICLR)},
  year      = {2024}
}

@inproceedings{distserve_zhong2024distserve,
  title     = {{DistServe}: Disaggregating Prefill and Decoding for Goodput-optimized Large Language Model Serving},
  author    = {Zhong, Yinmin and Liu, Shengyu and Chen, Junda and Hu, Jianbo and Zhu, Yibo and Liu, Xuanzhe and Jin, Xin and Zhang, Hao},
  booktitle = {18th USENIX Symposium on Operating Systems Design and Implementation (OSDI)},
  year      = {2024}
}

@inproceedings{amdahl1967validity,
  title     = {Validity of the Single Processor Approach to Achieving Large Scale Computing Capabilities},
  author    = {Amdahl, Gene M.},
  booktitle = {Proceedings of the AFIPS Spring Joint Computer Conference},
  pages     = {483--485},
  year      = {1967}
}

@misc{aws_fsx_lustre,
  title        = {{Amazon FSx for Lustre}},
  author       = {{Amazon Web Services}},
  howpublished = {\url{https://aws.amazon.com/fsx/lustre/}},
  year         = {2026},
  note         = {Accessed: 2026-10-08}
}

@misc{aws_s3,
  title        = {{Amazon Simple Storage Service (S3)}},
  author       = {{Amazon Web Services}},
  howpublished = {\url{https://aws.amazon.com/s3/}},
  year         = {2026},
  note         = {Accessed: 2026-10-08}
}

@misc{nvidia_gds,
  title        = {{NVIDIA GPUDirect Storage}},
  author       = {{NVIDIA}},
  howpublished = {\url{https://docs.nvidia.com/gpudirect-storage/}},
  year         = {2026},
  note         = {Accessed: 2026-10-08}
}

@misc{nvidia_nixl,
  title        = {{NIXL}: {NVIDIA} Inference Xfer Library},
  author       = {{NVIDIA}},
  howpublished = {\url{https://github.com/ai-dynamo/nixl}},
  year         = {2026},
  note         = {Accessed: 2026-10-08}
}

@inproceedings{gao2024cachedattention,
  title={Cost-Efficient Large Language Model Serving for Multi-turn Conversations with {CachedAttention}},
  author={Gao, Bin and He, Zhuomin and Sharma, Puru and Kang, Qingxuan and Jevdjic, Djordje and Deng, Junbo and Yang, Xingkun and Yu, Zhou and Zuo, Pengfei},
  booktitle={2024 USENIX Annual Technical Conference (USENIX ATC 24)},
  pages={111--126},
  year={2024}
}

@inproceedings{xiang2026servegen,
  title={{ServeGen}: Workload Characterization and Generation of Large Language Model Serving in Production},
  author={Xiang, Yuxing and Li, Xue and Qian, Kun and Zhang, Yan and Yu, Wenyuan and Zhai, Ennan and Jin, Xin and Zhou, Jingren},
  booktitle={23rd USENIX Symposium on Networked Systems Design and Implementation (NSDI 26)},
  pages={1845--1859},
  year={2026}
}

@article{rodriguez2002dynamic,
  title={Dynamic Parallel Access to Replicated Content in the {Internet}},
  author={Rodriguez, Pablo and Biersack, Ernst W.},
  journal={IEEE/ACM Transactions on Networking},
  volume={10},
  number={4},
  year={2002},
  doi={10.1109/TNET.2002.801413}
}

@article{bharadwaj2003divisible,
  title={Divisible Load Theory: A New Paradigm for Load Scheduling in Distributed Systems},
  author={Veeravalli, Bharadwaj and Ghose, Debasish and Robertazzi, Thomas G.},
  journal={Cluster Computing},
  volume={6},
  number={1},
  pages={7--17},
  year={2003}
}

@inproceedings{patterson1988raid,
  title={A Case for Redundant Arrays of Inexpensive Disks ({RAID})},
  author={Patterson, David A. and Gibson, Garth A. and Katz, Randy H.},
  booktitle={Proceedings of the 1988 ACM SIGMOD International Conference on Management of Data},
  pages={109--116},
  year={1988}
}

@misc{register2026rubin,
  author={{The Register}},
  title={Every Conference is an {AI} Conference as {Nvidia} Unpacks its {Vera Rubin} {CPUs} and {GPUs} at {CES}},
  year={2026},
  howpublished={\url{https://www.theregister.com/2026/01/05/ces_rubin_nvidia/}},
  note={Accessed: October 8, 2026}
}

\appendix

\section{Additional Experimental Results}
\label{sec:appendix}

This appendix backs claims of the evaluation, and the body cites each figure where it makes the claim: Appendix~\ref{sec:appendix:online} extends \S~\ref{sec:eval:online}, Appendix~\ref{sec:appendix:offline} extends \S~\ref{sec:eval:offline}, and Appendix~\ref{sec:appendix:demand} extends \S~\ref{sec:eval:envelope}. Setups follow \S~\ref{sec:eval:setup}: online runs serve the 32B model on one \polaris node, and offline runs serve LooGLE on one \jlse node, all with TP=4.

\subsection{\textbf{Online Serving}}
\label{sec:appendix:online}

\begin{figure*}[!ht]
    \minipage[t]{0.32\textwidth}
        \centering
        \includegraphics[width=\linewidth]{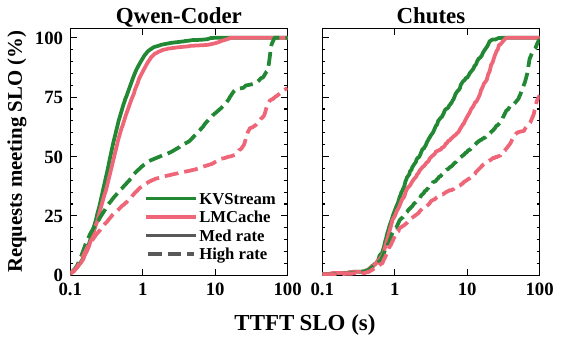}
        \captionsetup{skip=2pt,justification=raggedright}
        \caption{TTFT SLO attainment on the production traces at the medium and high arrival rates (Chutes at 40~GB of host memory per rank).}
        \label{fig:o9-slo}
    \endminipage
    \hfill
    \minipage[t]{0.32\textwidth}
        \centering
        \includegraphics[width=\linewidth]{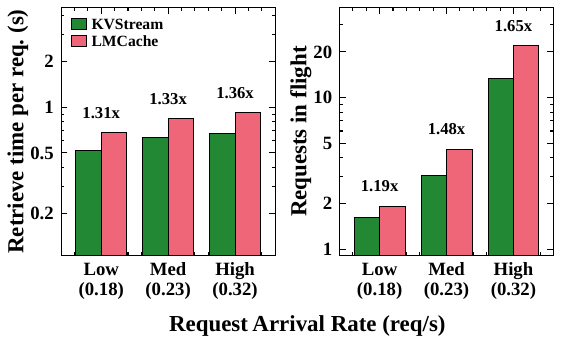}
        \captionsetup{skip=2pt,justification=raggedright}
        \caption{Retrieve time per request and requests in flight vs.\ arrival rate (Chutes, 40~GB of host memory per rank).}
        \label{fig:o10-chutes}
    \endminipage
    \hfill
    \minipage[t]{0.32\textwidth}
        \centering
        \includegraphics[width=\linewidth]{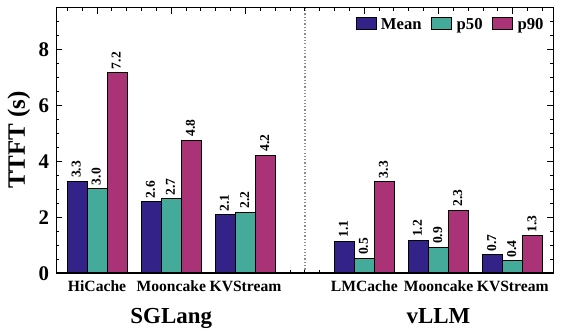}
        \captionsetup{skip=2pt,justification=raggedright}
        \caption{Mean, p50, and p90 TTFT with three storage backends per engine (SWE-bench agentic, first 3550 requests).}
        \label{fig:sglang}
    \endminipage
\end{figure*}

\paragraph*{\textbf{SLO attainment on the production traces}} Figure~\ref{fig:o9-slo} extends Figure~\ref{fig:o1-slo} to Qwen-Coder and Chutes at the medium and high arrival rates (Chutes with 40~GB of host memory per rank). At the medium rate, the two systems' curves nearly coincide: on Qwen-Coder, their median TTFTs differ by only 1.14$\times$, yet 3.35$\times$ fewer requests miss a 5~s SLO with \proj. At the high rate, more requests wait behind each retrieval, and \proj's curve lies well above LMCache's on both traces: the gain concentrates in the queued requests, as \S~\ref{sec:model:gains} predicts.

\paragraph*{\textbf{Retrieval versus queueing on Chutes}} Figure~\ref{fig:o10-chutes} repeats the decomposition of Figure~\ref{fig:o10-qwen} on Chutes. Per request, \proj retrieves about 1.3$\times$ faster at every rate, while the gap in requests in flight widens with the rate, so the queue, not the per-request saving, carries the speedup from 1.23$\times$ to 2.01$\times$ (\S~\ref{sec:model:gains}).

\paragraph*{\textbf{A second serving engine}} Figure~\ref{fig:sglang} repeats the SWE-bench comparison of Figure~\ref{fig:o1-slo} in SGLang and in vLLM, each with three storage backends. As HiCache's storage backend (\S~\ref{sec:impl}), \proj lowers SGLang's mean TTFT 1.57$\times$ against HiCache's file backend and 1.22$\times$ against Mooncake, although HiCache makes it write synchronously, without the deferred writes of M2. In vLLM, \proj has the lowest mean, p50, and p90 TTFT of the three backends, so the gain carries over to a second engine (MR-B.1).

\subsection{\textbf{Offline Throughput}}
\label{sec:appendix:offline}

\begin{figure*}[!ht]
    \minipage[t]{0.32\textwidth}
        \centering
        \includegraphics[width=\linewidth]{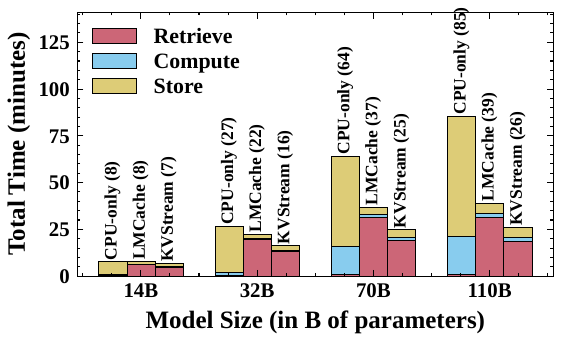}
        \captionsetup{skip=2pt,justification=raggedright}
        \caption{Processing-time breakdown across models, with \proj (LooGLE, \jlse).}
        \label{fig:model-breakdown}
    \endminipage
    \hfill
    \minipage[t]{0.32\textwidth}
        \centering
        \includegraphics[width=\linewidth]{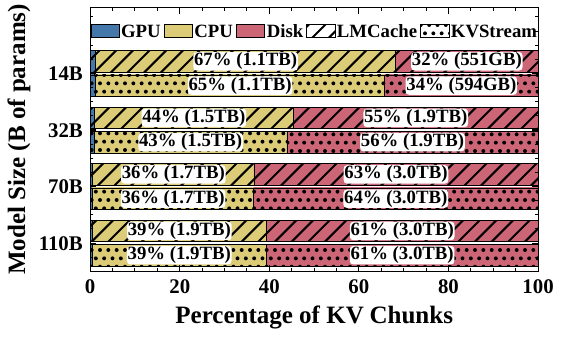}
        \captionsetup{skip=2pt,justification=raggedright}
        \caption{Share of KV chunks served from each tier across models (LooGLE, \jlse).}
        \label{fig:tier-dist}
    \endminipage
    \hfill
    \minipage[t]{0.32\textwidth}
        \centering
        \includegraphics[width=\linewidth]{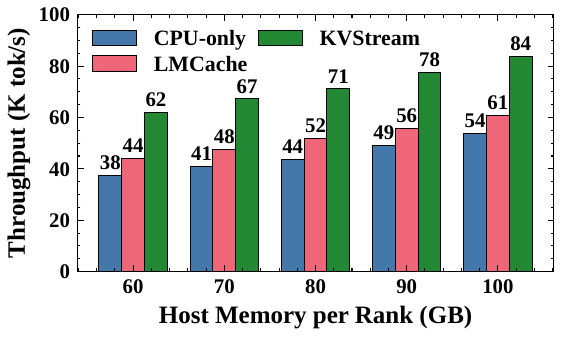}
        \captionsetup{skip=2pt,justification=raggedright}
        \caption{Throughput vs.\ host memory per rank (32B, LooGLE, \jlse).}
        \label{fig:cpu-memory}
    \endminipage
\end{figure*}

\paragraph*{\textbf{Where the time goes}} Figure~\ref{fig:model-breakdown} adds \proj to the breakdown of Figure~\ref{fig:g02-baselines} (O2) into the retrieve, compute, and store phases of Equation~\ref{eq:step-decomp}. LMCache and \proj spend nearly the same compute and store time at every model size, so the gain lies entirely in retrieval (\S~\ref{sec:eval:offline}): at 70B, \proj serves LooGLE in 25 minutes, against 37 for LMCache and 64 for CPU-only. Replication adds no visible store time offline, because each forward pass is long enough for the deferred writes (M2) to drain.

\paragraph*{\textbf{Same demand, faster reads}} Figure~\ref{fig:tier-dist} shows which tier serves the matched chunks. LMCache and \proj serve nearly the same shares from the GPU, host memory, and storage at every model size, so they face the same storage demand $D$ (Equation~\ref{eq:demand}), and \proj wins by reading it at the combined bandwidth of both tiers (Equation~\ref{eq:bound}, Figure~\ref{fig:io-throughput}). The storage share grows from about a third at 14B to about 60\% at 70B and 110B, the first of the two reasons why the gain grows with model size (\S~\ref{sec:eval:offline}).

\paragraph*{\textbf{Host memory offline}} Figure~\ref{fig:cpu-memory} raises host memory, the capacity $M$ of Equation~\ref{eq:demand}, from 60 to 100~GB per rank for the 32B model. Storage reads fall from 2,832 to 1,856~GB per rank and every approach speeds up, but LooGLE keeps spilling at every size, so \proj's advantage over LMCache stays at 1.37--1.41$\times$. Chutes, whose reuse distances straddle host memory, behaves differently (Appendix~\ref{sec:appendix:demand}).

\subsection{\textbf{Storage Demand and Write Cost}}
\label{sec:appendix:demand}

\begin{figure*}[!ht]
    \minipage[t]{0.32\textwidth}
        \centering
        \includegraphics[width=\linewidth]{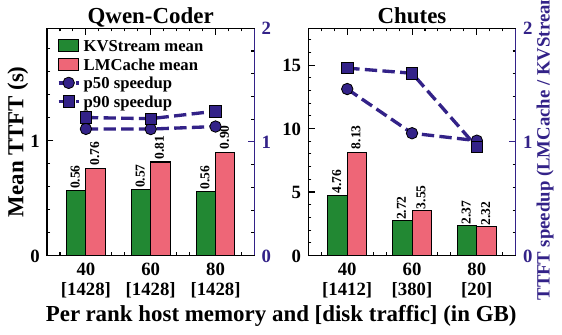}
        \captionsetup{skip=2pt,justification=raggedright}
        \caption{Mean TTFT and p50/p90 speedup vs.\ host memory at the medium arrival rates, with storage reads per rank in brackets.}
        \label{fig:o8-dram}
    \endminipage
    \hfill
    \minipage[t]{0.32\textwidth}
        \centering
        \includegraphics[width=\linewidth]{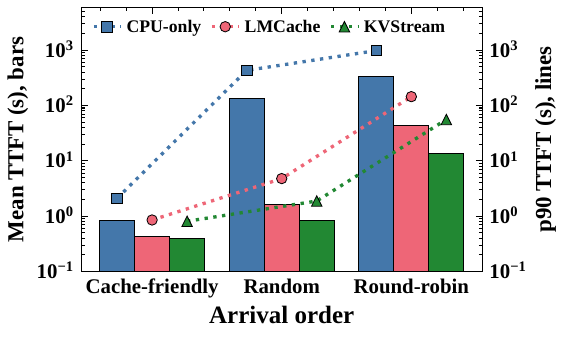}
        \captionsetup{skip=2pt,justification=raggedright}
        \caption{Mean (bars) and p90 (lines) TTFT under three arrival orders (Qwen-Coder, high arrival rate).}
        \label{fig:o4-order}
    \endminipage
    \hfill
    \minipage[t]{0.32\textwidth}
        \centering
        \includegraphics[width=\linewidth]{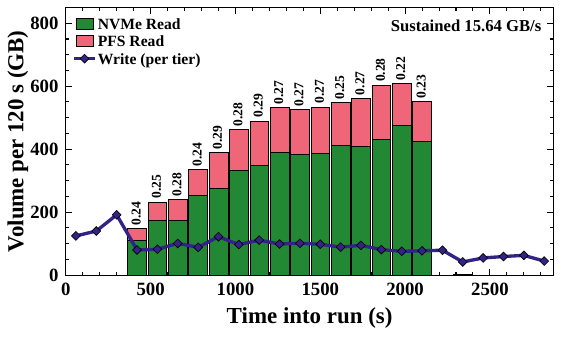}
        \captionsetup{skip=2pt,justification=raggedright}
        \caption{The round-robin run of Figure~\ref{fig:io-online}: storage reads by tier and writes per tier per 120~s, summed over four ranks.}
        \label{fig:o11-timeline}
    \endminipage
\end{figure*}

\paragraph*{\textbf{Host memory online}} Figure~\ref{fig:o8-dram} adds \proj to the sweep of Figure~\ref{fig:o5-memory} (O5), with the measured reads of Figure~\ref{fig:o12-reuse} in brackets. On Qwen-Coder, the reuse distances exceed every memory size (\S~\ref{sec:model:demand}): the reads stay at 1428~GB per rank, and \proj's mean TTFT stays near 0.56~s, against 0.76--0.90~s for LMCache. On Chutes, the reads fall from 1412 to 20~GB per rank and the speedup with them, from 1.71$\times$ to 0.98$\times$ (Figure~\ref{fig:reads-writes}): once memory holds the working set, the replicated writes have no reads to offset them (\S~\ref{sec:model:gains}).

\paragraph*{\textbf{Arrival order}} Figure~\ref{fig:o4-order} adds CPU-only and \proj to the order study of Figure~\ref{fig:o6-order-rate} (left, O6): Qwen-Coder at the high arrival rate, with 80~GB of host memory per rank. The order sets the reuse distances $d$ of Equation~\ref{eq:demand}, and the reads grow 5.8$\times$ from the cache-friendly to the round-robin order; \proj's speedup follows them, from 1.05$\times$ to 3.12$\times$ (Figure~\ref{fig:reads-writes}), with the lowest p90 in every order. CPU-only completes only 6200 and 4600 of the 6934 requests within the hour in the random and round-robin orders, so its bars there are lower bounds.

\paragraph*{\textbf{Replicated writes over time}} Figure~\ref{fig:o11-timeline} bins the round-robin run of Figure~\ref{fig:io-online} into 120~s intervals, summed over four ranks. Writes are exactly replicated (rank~0 writes 549.7~GB to each tier) and flow throughout, whereas reads begin once the cache fills and stop when the trace stops offering work. The PFS serves 22--29\% of each interval's reads, close to its 24\% share of the read bandwidth on \polaris (Table~\ref{tab:testbeds}): the split $1-\varphi$ of Equation~\ref{eq:phi}, which completion-driven routing (M4) reaches without measuring bandwidth.

\end{document}